\documentclass{aa}  

\usepackage{graphicx}
\usepackage{txfonts}
\usepackage{adjustbox}
\usepackage{amsmath}
\usepackage{amssymb}
\usepackage{natbib}
\usepackage{caption}
\usepackage{xcolor}
\usepackage[version=3]{mhchem}
\usepackage{xspace}
\usepackage{orcidlink}

\newcommand{\kms}{\ensuremath{\rm km\,s^{-1}}\xspace}
\newcommand{\Msun}{\ensuremath{\rm M_{\sun}}\xspace}

\newcommand{\Jyb}{\ensuremath{\rm Jy\,beam^{-1}}\xspace}
\newcommand{\vlsr}{\ensuremath{\rm v_{\mathrm{LSR}}}\xspace}
\newcommand{\draftone}[1]{\textcolor{black}{#1}}
\newcommand{\refereeone}[1]{\textcolor{black}{#1}}

\DeclareCaptionFormat{cont}{#1 (cont.)#2#3\par}

\begin{document}

   \title{Probing the Physics of Star-Formation (ProPStar)
   \thanks{Based on observations carried out under 
   project number S21AD with the IRAM NOEMA Interferometer and 
   090-21 with the IRAM 30m telescope. 
   IRAM is supported by INSU/CNRS (France), MPG (Germany) and IGN (Spain)}}
   \subtitle{II. The first systematic search for streamers toward protostars}
   
   \author{
        Mar\'ia Teresa Valdivia-Mena\inst{1}\orcidlink{0000-0002-0347-3837}
        \and
        Jaime E. Pineda\inst{1}\orcidlink{0000-0002-3972-1978}
        \and
        Paola Caselli\inst{1}\orcidlink{0000-0003-1481-7911}
        \and
        Dominique M. Segura-Cox\inst{2}\thanks{NSF Astronomy and Astrophysics Postdoctoral Fellow}\orcidlink{0000-0003-3172-6763}
        \and
        Anika Schmiedeke\inst{3}\orcidlink{0000-0002-1730-8832}
        \and
        Silvia Spezzano\inst{1}\orcidlink{0000-0002-6787-5245}
        \and
        Stella Offner\inst{2}\orcidlink{0000-0003-1252-9916}
        \and
        Alexei V. Ivlev\inst{1}\orcidlink{0000-0002-1590-1018}
        \and
        Michael Kuffmeier\inst{4}\orcidlink{0000-0002-6338-3577}
        \and
        Nichol Cunningham\inst{5,6}\orcidlink{0000-0003-3152-8564}
        \and
        Roberto Neri\inst{7}\orcidlink{0000-0002-7176-4046}
        \and
        Mar\'ia Jos\'e Maureira\inst{1}\orcidlink{0000-0002-7026-8163}
        }

   \institute{
   Max-Planck-Institut f\"ur extraterrestrische Physik, Giessenbachstrasse 1, D-85748 Garching, Germany\\
    \email{mvaldivi@mpe.mpg.de}
    \and
    Department of Astronomy, The University of Texas at Austin, 2500 Speedway, Austin, TX 78712, USA
    \and
    Green Bank Observatory, 155 Observatory Road, Green Bank, WV, 24944, USA
    \and
    Centre for Star and Planet Formation, Niels Bohr Institute, University of Copenhagen, Øster Voldgade 5-7, DK-1350
    Copenhagen, Denmark
     \and
    SKA Observatory, Jodrell Bank, Lower Withington, Macclesfield SK11 9FT, United Kingdom
    \and
    IPAG, Universit\'{e} Grenoble Alpes, CNRS, F-38000 Grenoble, France
    \and
    Institut de Radioastronomie Millimétrique (IRAM), 300 Rue de la Piscine, 38400 Saint-Martin-d’Hères, France
   }
   
   \date{}

 
  \abstract
   {The detection of \refereeone{narrow channels of accretion toward protostellar disks, known as streamers, have increased} in number in the last few years. However, it is unclear if streamers are a \draftone{common} feature around protostars that \refereeone{were previously} missed, or if they are a rare phenomenon.}
   {Our goals are to obtain the \draftone{incidence} of streamers toward a region of clustered star formation and to trace the origins of their gas, to determine if they originate \refereeone{within the} filamentary structure of molecular clouds \refereeone{or from beyond}. }
   {We used combined observations of the nearby NGC 1333 star-forming region, carried out with the NOEMA interferometer and the IRAM 30m single dish. Our observations cover the area between the IRAS 4 and SVS 13 systems. We traced the chemically fresh gas within NGC 1333 with \ce{HC3N} molecular gas emission and the structure of the fibers in this region with \ce{N2H^+} emission. We fit multiple velocity components in both maps and used clustering algorithms to recover velocity-coherent structures.}
   {We find streamer candidates toward 7 out of \draftone{16} \refereeone{young stellar objects} within our field of view. This represents \draftone{an incidence} of approximately \draftone{40\%} of \refereeone{young stellar objects} with streamer \draftone{candidates} when looking at a clustered star forming region. \refereeone{The incidence increases to about 60\% when we considered only embedded protostars}. All streamers are found in \ce{HC3N} emission.}
   {Given the different velocities between \ce{HC3N} and \ce{N2H^+} emission, and the fact that, by construction, \ce{N2H^+} traces the fiber structure, we suggest that the gas that forms the streamers comes from outside the fibers. This implies that streamers can connect cloud material that falls to the filaments with protostellar disk scales.}


   \maketitle
%

\section{Introduction}

Molecular clouds are composed of filaments \citep{Andre2014PPVIFilaments, Hacar2023PPVIIFilaments}, which contain the cores where protostars and binary systems are born \citep[e.g.][]{Andre2010Herschel, Offner2023PPVII}. \draftone{These filaments are highly dynamic structures: gas has been observed to flow along them and they also accrete more gas from their surroundings \citep[see][for more details]{Hacar2023PPVIIFilaments}.} 
Within filaments, molecular gas tends to be organized in velocity-coherent structures called fibers \citep[e.g.][]{Hacar2017NGC1333}.
Observations of filaments and fibers highlight their relevance in directing molecular gas toward the sites of star formation.

Although astronomers have observed and modeled how mass flows within filaments and fibers in general, how that mass reaches the protostellar disk (on scales \draftone{of $\lesssim0.1$ pc}) is not well understood. The way that mass reaches a protostellar system plays an important role in the star and planet formation process.
\draftone{For instance,} infall from the envelope to the disk that is variable in time influences the accretion rate of the protostar, affecting its luminosity \citep{Padoan2014Infallsim, Kuffmeier2018EpisodicAcc} and the chemical composition of the disk, represented, for instance, by the location of its snowline \citep{Hsieh2019-ALMA_Outbursts}
Numerical simulations also show that when infall from the envelope to the disk is heterogeneous in space, this can produce changes in disk structure, such as rings and gaps \citep{Kuznetsova2022asyminfall} and \refereeone{second-generation} disks \refereeone{misaligned with inner disks} \citep{Kuffmeier2021misaligneddisk}. 
Therefore, observing the mass flow from fiber scales ($\gtrsim 0.1$ pc or $\sim 20\,000$ au) to disk scales (few $\sim 100$ au) is crucial to understand the influence of filaments and fibers in protostellar and disk properties.

In the last few years there has been a rise in the observations of streamers, defined as velocity-coherent, narrow structures that \draftone{deposit their material to protostellar \refereeone{and protoplanetary} disks \citep{Pineda2023PPVII}}. They are observed as asymmetries in the protostellar envelopes with \draftone{total} lengths between 500 au \citep{Garufi2021accretionDGTauHLTau} to even 10 000 au away from the protostar, beyond the natal protostellar core \citep{Pineda2020Per2}. \refereeone{These structures are different from fibers, which are velocity-coherent structures within the filaments.} Streamers have been mostly detected and characterized toward embedded protostars \citep[which are known as Class 0 and I sources, e.g.][]{Chou2016DiskandFilconnectionL1455, Valdivia-Mena2022per50, Valdivia-Mena2023B5, Kido2023CB68streamer, Aso2023eDiskIRAS}, but some streamers have been detected toward T Tauri sources (also known as Class II protostars) as well \citep[e.g.][]{Ginski2021, Garufi2021accretionDGTauHLTau, Harada2023DKCha}. Although mostly characterized toward low-mass young stellar objects (YSOs), recently accretion streamers have been discovered toward high-mass YSOs as well \citep{Fernandez-Lopez2023massivediskstreamer}. 

\draftone{Although streamers appear increasingly common, their role in the larger puzzle of star formation is unclear.}. One open questions is where \draftone{do streamers} come from? Numerical simulations show asymmetric accretion channels \refereeone{potentially} generated by turbulence, both within the core itself  \citep[e.g.][]{Walch2010coresim, Seifried2013turb-mag-disks, Hennebelle2020disksim}, as well as coming from outside the natal core \citep{Kuffmeier2017zoom-insims,Kuffmeier2023rejuvenatinginfall, Heigl2024streamerform}. \refereeone{Observations of different species of molecular gas suggest that streamers can transport gas that comes from beyond the filaments themselves. For instance,} \cite{Valdivia-Mena2023B5} suggest that the observed gas falling toward fibers at scales of $\sim 20\,000$ au is connected to a streamer feeding disk scales in Barnard 5, but \draftone{they} cannot \draftone{directly confirm this suggestion} due to the different tracers and \draftone{spatial} resolutions of the observations. 

Another point of debate is how frequent streamers actually are. The previously mentioned discoveries of streamers are mostly serendipitous. It is possible that accretion streamers are a common feature within protostellar envelopes of all ages, but they have not been characterized because the \refereeone{observations} targeted other features, such as the disk or outflows \citep[e.g.][]{Thieme2022Lupus3streamers}. For instance, the \draftone{the primary goal of} the ALMA Large Program eDisk \draftone{is} to find substructures in young, embedded disks, \draftone{yet} their observations reveal streamers as well \citep{Kido2023CB68streamer, Aso2023eDiskIRAS}. To this date, there has not been a systematic search for streamers toward YSOs. To better understand the role of streamers in our new picture of star formation, it is necessary to actively search for streamers in data that can also trace the larger fiber kinematics.

\draftone{We} present the first systematic study of gas flow from filament scales to individual protostellar envelopes with the explicit goal to search for streamers. \draftone{This is part of the ``Probing the Physics of Star-formation" (ProPStar) survey \citep{Pineda2024ProPStarI}, where we explore the connection between the molecular gas within filaments and the circumstellar disk scales.} We trace the flow of gas from a filament to \refereeone{YSOs}, using a set of observations that allows us to characterize the flow of gas at fiber scales as well as toward individual protostars. The goal of this work is to search for streamers in a systematic fashion \draftone{within} a region \draftone{where the kinematic properties of the fibers are known}, to investigate the connection between fibers and streamers. For this purpose, we selected NGC 1333, one of the closest young embedded clusters, containing close to 150 YSOs \citep{Gutermuth2008spitzerNGC1333}. It is the most active star forming region within the Perseus molecular cloud, located at 293 pc from Earth \citep{Ortiz-Leon2018perseusdist, Zucker2018distance}. Its high protostellar activity is reflected in the dozens of outflows stirring the local gas \citep{Plunkett2013outflows}. \draftone{For this work, we select a region that covers two fibers with known kinematic properties which include the SVS 13 and the IRAS 4 protostellar systems. The observed area includes a total of 16 \refereeone{YSOs} between Class 0 and Class II,} \refereeone{enabling a first pilot study of incidence of streamers within an active star forming region.} We refer to this whole area as the southeast (SE) filament.

This article is divided as follows: In Section \ref{sec:datared}, we describe the observations with the NOEMA and 30-m telescopes and their combination. In Section \ref{sec:methods}, we describe the \refereeone{methods used to decompose the observed line emission into velocity-coherent clusters for kinematic analyses.} Section \ref{sec:results} shows the resulting velocity structure of the fibers seen in the chosen gas tracers, the steps followed to find streamers in the data, and a description of the \refereeone{found candidates}. We discuss our results and their \draftone{physical} interpretation in Section \ref{sec:discussion}. Finally, in Section \ref{sec:conclusions}, we summarize the main results of our work. 

\section{Observations and Data reduction\label{sec:datared}}

We summarize the single dish and interferometric observations of NGC 1333 SE that we use from the ProPStar survey. The same observational setup is described in the ProPStar I paper \citep{Pineda2024ProPStarI}. 

\begin{figure*}
    \centering
    \includegraphics[width=0.48\textwidth]{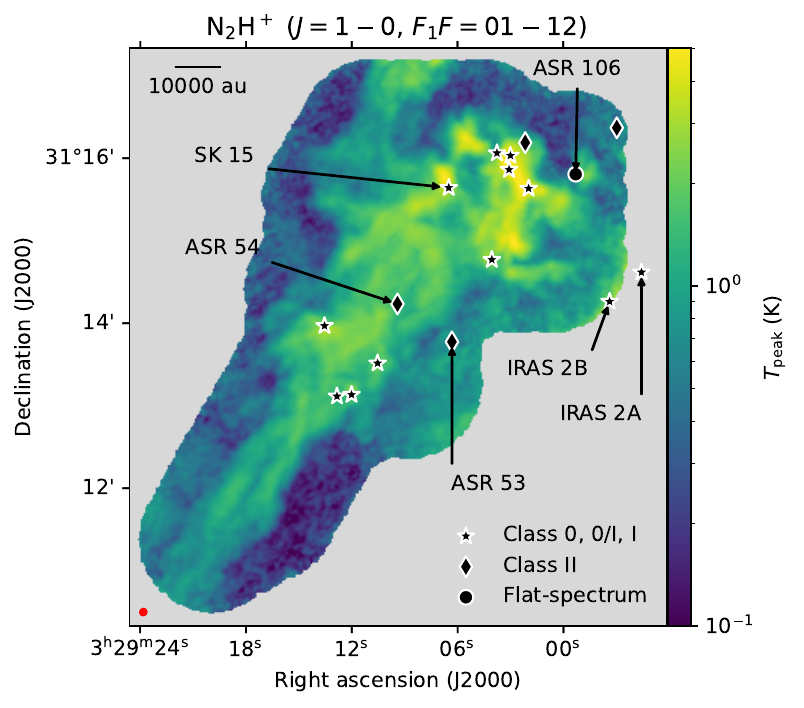}
    \includegraphics[width=0.48\textwidth]{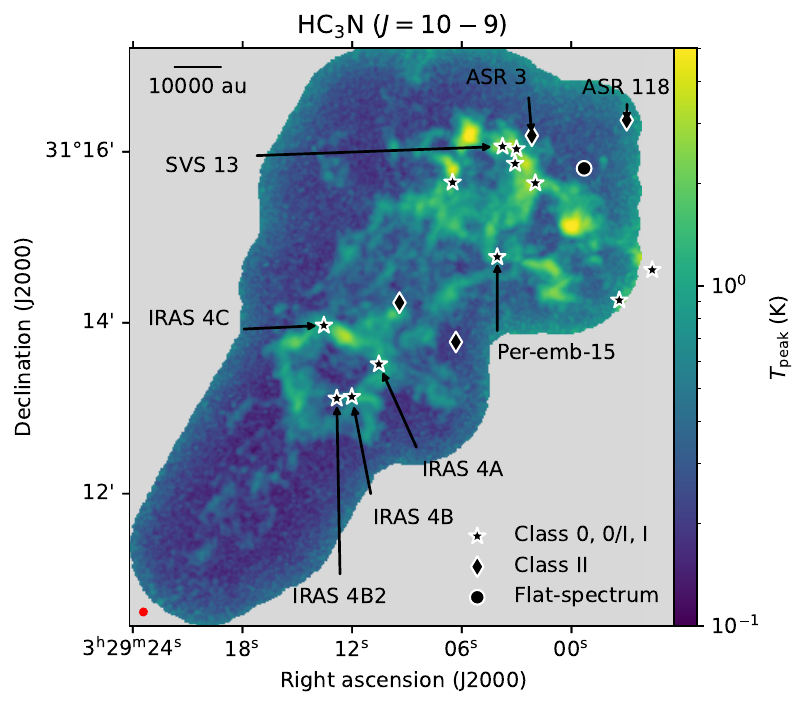}
    \caption{\draftone{Peak temperature $T_{\mathrm{peak}}$ maps of the NOEMA and 30-m telescope observations. The black symbols represent \refereeone{YSOs} in the region, summarised in Table \ref{tab:protostars}: stars mark the positions of Class 0, I and 0/I protostars, circles represent Flat-spectra objects, and diamonds mark Class II \refereeone{sources}. The protostars are labeled following Table \ref{tab:protostars}.}  \refereeone{Left: \ce{N2H^+} ($1-0$) $F_1F = 01-12$ $T_{\mathrm{peak}}$. Right: \ce{HC3N} ($10-9$) $T_{\mathrm{peak}}$. }} 
    \label{fig:mom8maps}
\end{figure*}

\subsection{IRAM 30-m telescope}

The single-dish observations were carried out using the 30-m telescope from IRAM, located in the Pico Veleta \draftone{mountain in} Spain. The observations are part of project 091-21 and were done during 2021 November 9, 10, and 11; and 2022 February 19, 20. We tuned the EMIR 090 receiver to cover the ranges between 72.3 – 78.8 GHz and 87.8 – 94.3 GHz. The FTS50 backend was employed, \draftone{yielding a spectral resolution of 52.5 kHz}. The complete map is $\approx150\arcsec\times150\arcsec$ and was \refereeone{observed} using an on-the-fly technique with position switching. 

Data reduction was performed using the CLASS program of the GILDAS package\footnote{\url{http://www.iram.fr/IRAMFR/GILDAS}}. 
The beam efficiency, $B_{eff}$, was obtained using the Ruze formula (available in CLASS), and was used to convert the observations into main beam temperatures, $T_{\rm mb}$.

\subsection{NOEMA}

The observations were carried out with the IRAM NOrthern Extended Millimeter Array (NOEMA) interferometer as part of the the S21AD program using the Band 1 receiver, \draftone{with the array in} D configuration. \draftone{The observations were taken} on 2021 July 18, 19, and 21; August 10, 14, 15, 19, 22, and 29; and September 1. The mosaic consists on 96 pointings, centered at $\alpha_{J2000}$=03$^{\rm h}$29$^{\rm m}$10.2$^{\rm s}$, $\delta_{J2000}$=31$\degr$13$\arcmin$49.4$\arcsec$, which were \refereeone{observed} on four different scheduling blocks. We used the PolyFix correlator with a LO frequency of 82.505\,GHz, an instantaneous bandwidth of 31 GHz spread over two sidebands (upper and lower) and two polarisations.
The centers of the two 7.744\,GHz wide sidebands are separated by 15.488\,GHz. 
Each sideband is composed of two adjacent basebands of $\sim$3.9\,GHz width (inner and outer basebands). In total, there are eight basebands which were fed into the correlator. The spectral resolution is 2\,MHz throughout the 15.488\,GHz effective bandwidth per polarization. Additionally, a total of 112 high-resolution chunks are placed, each with a width of 64\,MHz and a fixed spectral resolution of 62.5\,kHz. Both polarizations (H and P) are covered with the same spectral setup, and therefore the high-resolution chunks provide 66 dual polarisation spectral windows. In this work, we used the spectral windows containing the \ce{N2H^+} $J=1-0$ line \draftone{and its hyperfine components ($F_1F$)} \refereeone{at approximately 93 GHz}, and the \ce{HC3N} $J=10-9$ line \refereeone{at approximately 91 GHz (Table \ref{tab:lines})}.

Data reduction was done using the CLIC program from the GILDAS package. We used the NOEMA pipeline to obtain the calibrated uv-tables, which we then combined with the single-dish data as explained in the next section.

\subsection{Combination of single-dish and interferometric data}

\begin{table*}[htbp]
\caption{Spectral lines observed in the high-resolution chunks used in this work\label{tab:lines}}
\centering
\begin{tabular}{ccccccc}
 \hline \hline
Molecule & Transition & $\nu$ (MHz) & \draftone{E$_{\mathrm{up}}$ (K)} & \draftone{$n_{\mathrm{crit}}$ (cm$^{-3}$) }& beam size   & rms (mK)\\
 \hline
\ce{N2H^+} &  $J=1-0$, \draftone{$F_1F = 01-12$}         &   \draftone{93176.2522} & \draftone{4.47} & $6.1\times 10^{4}$ & $5.76 \times 5.48$ & 70 \\
\ce{HC3N} & $J=10-9$       & 90979.0230 & \draftone{24.01}  & $1.6\times 10^{5}$ & $4.89 \times 4.69$ & 80  \\
 \hline
\end{tabular}
\tablefoot{Frequencies taken from the CDMS catalog. \draftone{Critical densities taken from \cite{Shirley2015ncrit}.} The rms is taken from the combined NOEMA and 30-m data cubes.} 
\end{table*}

\refereeone{The combination and imaging of the datasets previously described was done in \texttt{mapping}.} The original 30-m data is resampled to match the spectral setup of the NOEMA observations. 
We use the task \verb+uvshort+ to generate the pseudo-visibilities 
from the 30-m data for each NOEMA pointing. 
The imaging is done with natural weighting, a support mask, and 
using the SDI deconvolution algorithm.

The final cubes are in K and have a channel resolution of approximately 0.2\,\kms. The properties of the final data cubes are described in Table \ref{tab:lines}. The channel maps for each of the molecules are \draftone{presented} in Appendix \ref{ap:chanmaps}. The observations cover an area of approximately $150\arcsec \times 150\arcsec$. As we are interested in the kinematic structure of the filaments, we isolated one of the hyperfine structure emission lines of \ce{N2H^+} $J=1-0$, the $F_1F = 01-12$ component at 93176.2522\,MHz. 
We chose this \draftone{hyperfine} component because it is the one that is the most separated from the other hyperfine structure lines, and therefore each individual peak can be interpreted as an individual velocity structure along the line of sight \citep[Fig. 1 in][]{Caselli1995N2Hp}. The \ce{N2H^+} $J=1-0$, $F_1F = 01-12$ \draftone{peak temperature} map is in Fig.~\ref{fig:mom8maps} \refereeone{left}. The clean beam size for this molecule is approximately 5.8\arcsec (1750\,au). The mean rms of \ce{N2H^+} ($1_{01}-0_{12}$) cube is approximately 15.7 m\Jyb (70\,mK). \draftone{For the rest of the article, we refer to the \ce{N2H^+} $J=1-0$, $F_1F = 01-12$ as \ce{N2H^+} unless otherwise stated.}

The \ce{HC3N} $J=10-9$ \draftone{peak temperature} map is shown in Fig.~\ref{fig:mom8maps} \refereeone{right}. The clean beam size at this frequency is approximately 4.9\arcsec (1400\,au). The mean rms of the \ce{HC3N} $J=10-9$ emission cube is approximately 12\,m\Jyb (80\,mK) and increases at the edges of the map due to the primary beam response. \draftone{We refer to \ce{HC3N} ($10-9$) line emission as \ce{HC3N} in the rest of this work, as is the only \ce{HC3N} emission line presented.}

\begin{table*}[htbp]
\centering
\caption{\label{tab:protostars}Properties of the protostellar objects found within the  \ce{N2H^+} and \ce{HC3N} maps.}
\adjustbox{max width=\textwidth}{
\begin{tabular}{ccccccccc}
\hline \hline
Source                     & RA  & DEC  & Other Names                               & Class$^{a}$ & Multiple? $^{b}$ & $\vlsr$$^c$ & Outflow P.A.$^d$ & Disk $i^e$ \\ 
 & (J2000) & (J2000) & & & &  (\kms) & ($^{\circ}$)  & ($^{\circ}$)  \\
\hline
IRAS4A                     & 03:29:10.54            & +31:13:30.93       & Per-emb 12                                & 0     & Y                & $6.9 \pm 0.005^1$               & 19$^6$                 & 35                         \\
IRAS4B                     & 03:29:12.02            & +31:13:08.03       & Per-emb 13                                & 0     & Y                & $7.1 \pm 0.009^1$               & 176$^6$                & 49                         \\
Per-emb-14                 & 03:29:13.55            & +31:13:58.15       & NGC 1333 IRAS4C                           & 0     & N                & $7.9 \pm 0.03^1$                & 96$^7$                 & 64                         \\
Per-emb-15                 & 03:29:04.06            & +31:14:46.24       & RNO15-FIR, SK 14                          & 0     & N                & $6.8 \pm 0.01^1$                & -35$^5$                &                            \\
Per-emb-27                 & 03:28:55.57            & +31:14:37.03       & NGC 1333 IRAS2A                           & 0/I   & Y                & $8.1 \pm 0.02^1$                & 105$^{5,f}$            &                            \\
Per-emb-36                 & 03:28:57.37            & +31:14:15.77      & NGC 1333 IRAS2B                           & I     & Y                & $6.9 \pm 0.02^1$                & 24$^5$                 &                            \\
Per-emb-44                 & 03:29:03.76            & +31:16:03.81       & SVS13A                                    & 0/I   & Y                & $8.7 \pm 0.02^1$                & 140$^8$                &                            \\
SVS13C                     & 03:29:01.97            & +31:15:38.05       &                                           & 0     & Y                & $8.9 \pm 0.02^1$                & 8$^8$                  & 75                         \\
SVS13B                     & 03:29:03.08            & +31:15:51.74       &                                           & 0     & Y                & $8.5 \pm 0.01^1$                & 170$^6$                & 61                         \\
IRAS4B2                    & 03:29:12.84            & +31:13:06.89       & NGC 1333 IRAS4B'                          & 0     & Y                &                                 &          -99$^9$           &                            \\
EDJ2009-183                & 03:28:59.30            & +31:15:48.41       & ASR 106                                   & Flat  & Y                & $8.69^2$                        &                        &                            \\
EDJ2009-173                & 03:28:56.96            & +31:16:22.20       & ASR 118, SVS15                            & II    & N                & $9.08^2$                        &                        &                            \\
VLA 3                      & 03:29:03.00             & +31:16:02.00        &                                           & 0     &                &                                 &                        &                            \\
SK 15                      & 03:29:06.50             & +31:15:38.60        & ASR 6, HRF 50                             & I     &                & $7.95^3$                        &                        &                            \\
ASR 3                      & 03:29:02.16             & +31:16:11.40        & {[}GMM2008{]} 76 & II    &                &                                 &                        &                            \\
ASR 53 & 03:29:02.16             & +31:16:11.40        & {[}GMM2008{]} 89                  & II    &                & $13.5 \pm 3.4^4$                &                        &                            \\
ASR 54 & 03:29:09.41             & +31:14:14.10        & {[}GMM2008{]} 135                 & II    &                &                                 &                        &                            \\ 
\hline
\end{tabular}
}
\tablefoot{We only describe the properties of protostars that are on regions where there is \ce{HC3N} emission or the outflow lobe is within the \ce{HC3N} emission map.
\tablefoottext{a}{SED classification from the literature \citep{Enoch2009classification, Evans2009c2d}. We indicate also when a source is a Class 0/I according to the classification in the VANDAM sample \citep{Tobin2018VANDAMbinaries}.}
\tablefoottext{b}{Y if the source contains more than one protostar, N if it is single \citep{Tobin2018VANDAMbinaries}}
\tablefoottext{c}{Obtained from available molecular tracer observations of cores or from spectroscopic surveys. References: (1) \cite{Stephens2019MASSES}, (2) \cite{Foster2015virialcores-prems}, (3) \cite{Imai2018deuterium}, (4) \cite{Kounkel2019closecompanions}.}
\tablefoottext{d}{Angle of the red lobe from North to East. References: (5) \cite{Stephens2017fil-out-alignment}, (6) \cite{Lee2016MASSES}, (7) \cite{Zhang2018IRAS4Coutflow}, (8) \cite{Plunkett2013outflows}, (9) \cite{Podio2021calypso}}
\tablefoottext{e}{Disk inclination with respect to the plane of the sky from \cite{Segura-Cox2018AVANDAM}.} 
\tablefoottext{f}{In IRAS 2A, there are two outflows perpendicular to each other. Here we list the most collimated one, where the red lobe is included in the footprint of our \ce{HC3N} data. }
}
\end{table*}

The \refereeone{YSOs} \draftone{that are found} within the \draftone{observed area} and their general characteristics are summarised in Table \ref{tab:protostars}.  \draftone{We plot these protostars and \refereeone{Class II sources} in Fig.~\ref{fig:mom8maps} and label each one so that the reader has a quick reference of where each one is in for the following analyses. We include IRAS 2A even though our map does not cover the protostar as it has a strong outflow that needs to be taken into account to understand the gas kinematics of the region}.

\section{\refereeone{Line decomposition methods}\label{sec:methods}}

The peak temperature maps shown in Fig. \ref{fig:mom8maps} do not fully reflect the complexity of the obtained data. Both molecular lines show several emission peaks, some located at or within a beam of protostars, others apparently unrelated to protostellar sources. However, an initial inspection of the data cube shows that there are several locations in the map where the spectra have two or three velocity components along the line of sight. Previous molecular gas observations done both with similar resolution \refereeone{to our data} \citep{Dhabal2019NGC1333NH3} as well as with single-dish telescopes \citep{Hacar2017NGC1333,Dhabal018serpens-perseus, Chen2020b-velstructureNGC1333} show that there are multiple velocity components in this region. \refereeone{In particular,} \cite{Hacar2017NGC1333} and \cite{Chen2020b-velstructureNGC1333} find that these velocity components are due to three velocity-coherent fibers of gas in the region, two of which run in the northwest to southeast direction that overlap in line of sight, \refereeone{although the exact structure identification is different for both works (we discuss the differences further in Appendix \ref{ap:fiber-comparison}).} For the kinematic analysis of the gas surrounding the protostars, it is necessary to \draftone{separate these velocity components.}

In this section, we \draftone{describe the procedure used to separate the velocity components} of NGC 1333 SE. \draftone{In summary,} we \draftone{first} separated the individual velocity components of each spectrum within the \ce{N2H^+} and \ce{HC3N} maps by fitting multiple Gaussians. \draftone{Next}, we clustered the velocity components to recover individual, \draftone{velocity-coherent} fibers in the region. \draftone{This process allowed us to analyze the velocity structure around the protostars more easily.} 

\subsection{Identification of velocity components\label{sec:bayesian}}

Fitting a model using a simple minimization algorithm, such as least squares, has issues in identifying the best parameters for faint signals. Moreover, quantifying how much better a more complex model (i.e. with more parameters) is compared to a simpler one is a delicate task \citep[e.g.][]{Protassov2002}. A quick look at the data cubes indicates that there are \draftone{lines of sight} where the possible second component has a signal-to-noise ratio (S/N) $\sim5$. Therefore, a simple least squares fit followed by a statistic evaluation might not be enough to effectively separate the kinematic components within the filaments.

Bayesian model selection using nested sampling can solve these problems. Nested sampling is a parameter exploration algorithm that evaluates how probable the combination of parameters is, therefore allowing parameter estimation and model comparison, even in low S/N data \citep{Skilling2004NS, Skilling2006NS,Sokolov2020BayesFit}. \refereeone{This} algorithm returns the Bayesian Evidence $\mathcal{Z}$ of a model, which is a likelihood integral of the parameter values:

\begin{equation}
    \mathcal{Z} = \int p(\theta) \mathcal{L}(\theta) d\theta,
\end{equation}
where $\theta$ are the probable parameter values, $p(\theta)$ is the probability density function, and $\mathcal{L}(\theta)$ is the likelihood function. From $\mathcal{Z}$, we can then calculate the relative probability $K_{n-1}^{n}$ that a number of components $n$ \draftone{returns a better fit} than $n-1$ components:

\begin{equation}
    K_{n-1}^{n} = \frac{P(\mathcal{M}_{n})\mathcal{Z}_{n}}{P(\mathcal{M}_{n-1})\mathcal{Z}_{n-1}} = \frac{\mathcal{Z}_{n}}{\mathcal{Z}_{n-1}},
\end{equation}
where $P(\mathcal{M}_{n})$ is the probability \textit{a priori} of a model with $n$ Gaussian components (for our case). We assume that all competing models are equally likely \textit{a priori}, so $P(\mathcal{M}_{n}) = P(\mathcal{M}_{n-1})$.

\begin{figure*}[htb]
	\centering
	\includegraphics[width=0.95\textwidth]{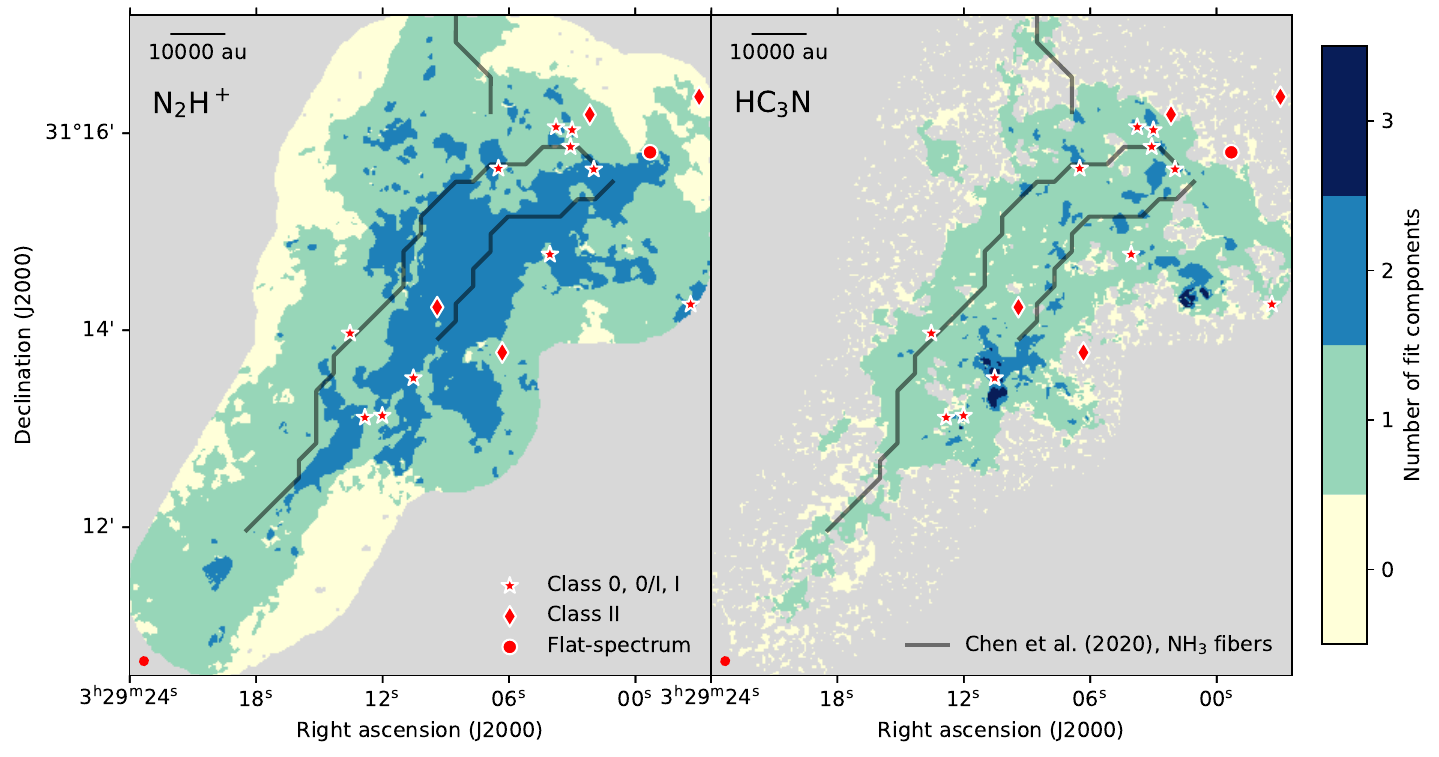}
	\caption{Resulting number of \refereeone{individual velocity} components fitted along each line of sight for \ce{N2H^+} (left) and \ce{HC3N} (right). The red symbols represent protostellar objects in the region as in Fig. \ref{fig:mom8maps}. \draftone{The gray regions correspond to pixels with SNR$\leq2$ or areas without data.} \refereeone{The black semitransparent lines show the spines of the fibers in \ce{NH3} \citep{Chen2020b-velstructureNGC1333}.} The red ellipse at the bottom left corner \draftone{of each plot} represents the beam \draftone{size}. }
	\label{fig:res-ncomp}
\end{figure*}

We used the \texttt{gaussian} model in the Python package \texttt{pyspeckit} to fit the spectra with 1 and 2 Gaussian components for \ce{N2H^+} and up to 3 Gaussian components for \ce{HC3N}. The third component in the case of \ce{HC3N} is to account for outflow emission. We used \texttt{pymultinest} \citep{Buchner2014X-raypymultinest}, a package designed to use the Fortran-based package \texttt{MultiNest} in Python \citep{Feroz2009Multinest, Feroz-Hobson2008Multinestmethod}, to run the nested sampling.
In conjunction to \texttt{pymultinest}, we used the Python package 
\texttt{pyspecnest}\footnote{https://github.com/vlas-sokolov/pyspecnest} \citep{Sokolov2020BayesFit} to wrap the \texttt{pyspeckit} fitting classes so as to use them in \texttt{pymultinest}. We followed the steps and code used in \cite{Sokolov2020BayesFit}\footnote{https://github.com/vlas-sokolov/bayesian-ngc1333}. 

From this process, we obtained the best fit parameters and their uncertainties for every spectra in the cubes with SNR$ > 2$. We adopted a conventional cut \draftone{for $K$ of} $\ln  K_{n-1}^{n} > 5$, so if $\ln K_{n-1}^{n} \leq 5$, we \draftone{select} $n-1$ components. We use this criterion to \draftone{obtain a map with the number of components $n$ in each pixel. There are small regions (consisting of less than 100 pixels) of emission in this map that are surrounded by noisy data, so we ran two additional steps to reduce the variation of $n$ due to noise. } 

First, we eliminated small regions of consisting of \draftone{$n$ components surrounded by $n-1$ components or emission surrounded by noise ($n=0$), which we call islands}, \refereeone{by replacing them with their resulting $n-1$ fit results. Islands of less than 100, 7 and 2 pixels for the cases of $n=1$,  $n=2$ and $n=3$ respectively were replaced with the results from $n-1$ fit.} These thresholds are based on a visual inspection of the resulting fits, where islands with a size smaller than the threshold tend to fit noise peaks outside the observed emission range (5 to 10 \kms). Nevertheless, the exact \draftone{size of the islands} does not influence the final results. 

\draftone{\refereeone{Secondly, we filtered} the fit results using the parameter uncertainties. For our analysis,} we only used the fit results that have an uncertainty in central velocity smaller than the channel size. We evaluated each individual fit component to determine its quality. If the central velocity of one component has a larger uncertainty than the channel width, that component was eliminated, without affecting the other components that might be present in the same spectrum. The $n$ map is updated accordingly, for instance, if $n=3$ in a pixel and one component was eliminated, we replaced $n=2$ in that position. After this filtering, we obtained central velocities that have uncertainties between 0.02 and 0.03 \kms for \ce{N2H^+}, and between 0.03 and 0.05 \kms in for \ce{HC3N} . 

\draftone{Figure \ref{fig:res-ncomp} shows the final number of Gaussian components per spectrum in the \ce{N2H^+} (left) and \ce{HC3N} (right) emission cubes.} We required up to 2 components for \ce{N2H^+} and up to 3 in \ce{HC3N} to adequately fit the spectra, \draftone{where the} third Gaussian component in all cases was used to fit extended wings of emission in the \ce{HC3N} spectra. \draftone{Section \ref{sec:hc3n-outflows} describes these  \refereeone{wings} in more detail.} 

\begin{figure*}
	\centering
	\includegraphics[width=0.9\textwidth]{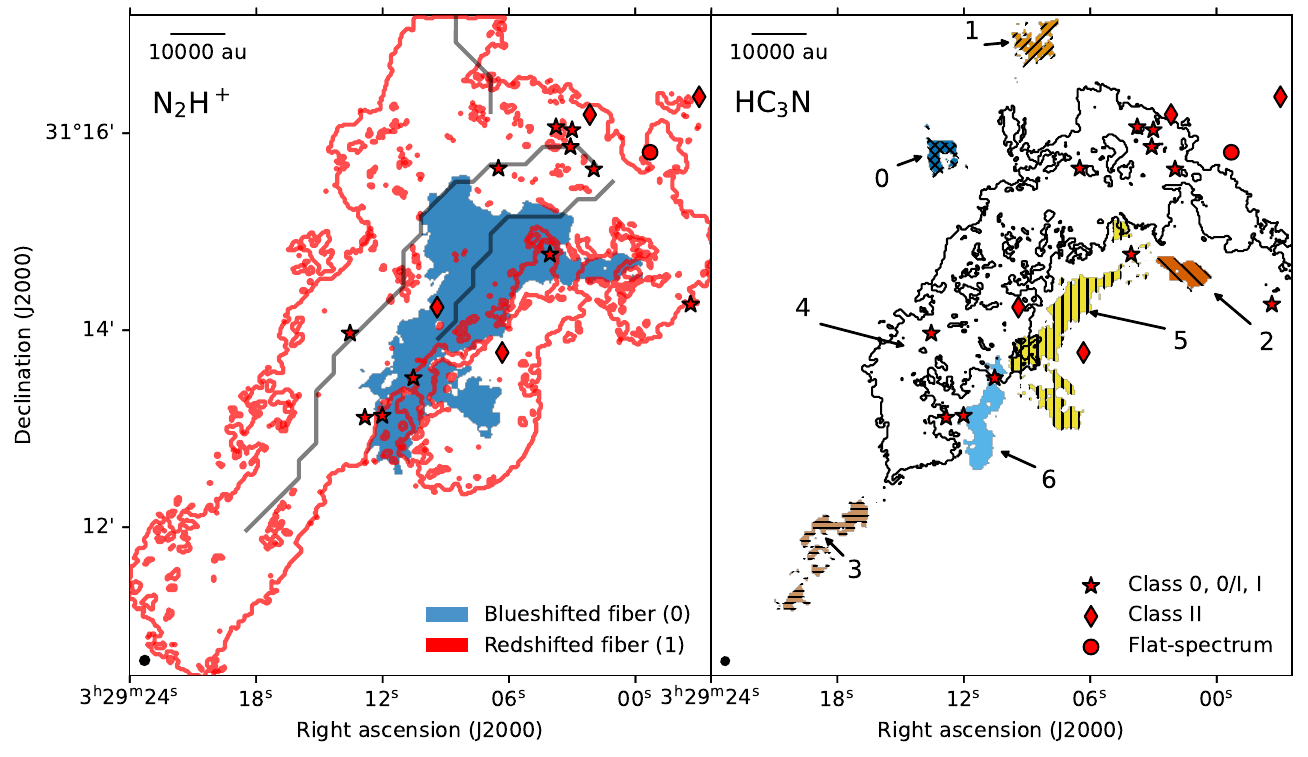}
	\caption{Results of the clustering algorithms for the Gaussian components of   \refereeone{\ce{N2H^+} and \ce{HC3N}}. The black circles at the bottom left corners represent the beam size of the respective data. The scalebar at the top left corner represents a length of 10 000 au.  \refereeone{The black semitransparent lines show the spines of the fibers in \ce{NH3} \citep{Chen2020b-velstructureNGC1333}.} \draftone{Left: cluster groups for \ce{N2H^+}, labeled as red and blue representing the more redshifted and blueshifted groups, respectively. Right: cluster groups for \ce{HC3N}, labeled 0 to 6.}}
	\label{fig:cluster-results}
\end{figure*}

\subsection{Clustering of velocity structures\label{sec:clustering}}

\draftone{We clustered the individual velocity components of each of the molecular gas emissions to compare \ce{HC3N} emission with \ce{N2H^+} emission. This comparison is not straightforward due to the presence of multiple velocity components distributed differently in both molecules. Figure \ref{fig:res-ncomp} \refereeone{left} shows that \ce{N2H^+} has mostly two velocity components at the center of the map, \refereeone{located similarly to the overlapping fibers found with lower resolution data \ce{NH3} (approximately 30\arcsec) by \cite{Chen2020b-velstructureNGC1333}.} However, \ce{HC3N} does not show this overlap, with regions of two and three velocity components scattered throughout the map \refereeone{(Fig. \ref{fig:res-ncomp} right)}. Therefore, it is not clear which \refereeone{Gaussian} component in \ce{HC3N} should be compared with what \ce{N2H^+} component. By clustering the Gaussian peaks, the comparison between both molecules is possible.}

\draftone{We clustered the components using a density-based clustering algorithm because it is not possible to separate them manually using a simple velocity threshold. If we grouped all velocity components larger than a certain \vlsr together, it is possible we separated emission that is connected with the grouped emission but that has a large velocity gradient, and so has components with a lower \vlsr than the threshold. }
\draftone{A look at the velocities obtained after the nested sampling (Fig.  \ref{fig:N2Hp-clusterprops} and \ref{fig:HC3N-clusterprops}) shows that both molecules present complex velocity structures.}

We describe all the details of the clustering process for each molecule in Appendix \ref{ap:clustering}. \draftone{In summary, we used Hierarchical Density-Based Spatial Clustering of Applications with Noise (HDBSCAN) to find clusters of emission in position-position-velocity space. We excluded from the clustering the Gaussian components with $\sigma_{\mathrm{v}}>1$ \kms, as these trace outflow cones in \ce{HC3N} (Sect. \ref{sec:hc3n-outflows}). Figure \ref{fig:cluster-results} shows the resulting clusters and their labels.} 
The complete set of properties (peak temperature, velocity and velocity dispersion) are in Fig. \ref{fig:HC3N-clusterprops} for \ce{HC3N} and in Fig. \ref{fig:N2Hp-clusterprops} for \ce{N2H^+} clusters.

\section{Results and analysis\label{sec:results}}

\subsection{Properties of the NGC 1333 SE gas}

\begin{figure*}
	\centering
	\includegraphics[width=0.94\textwidth]{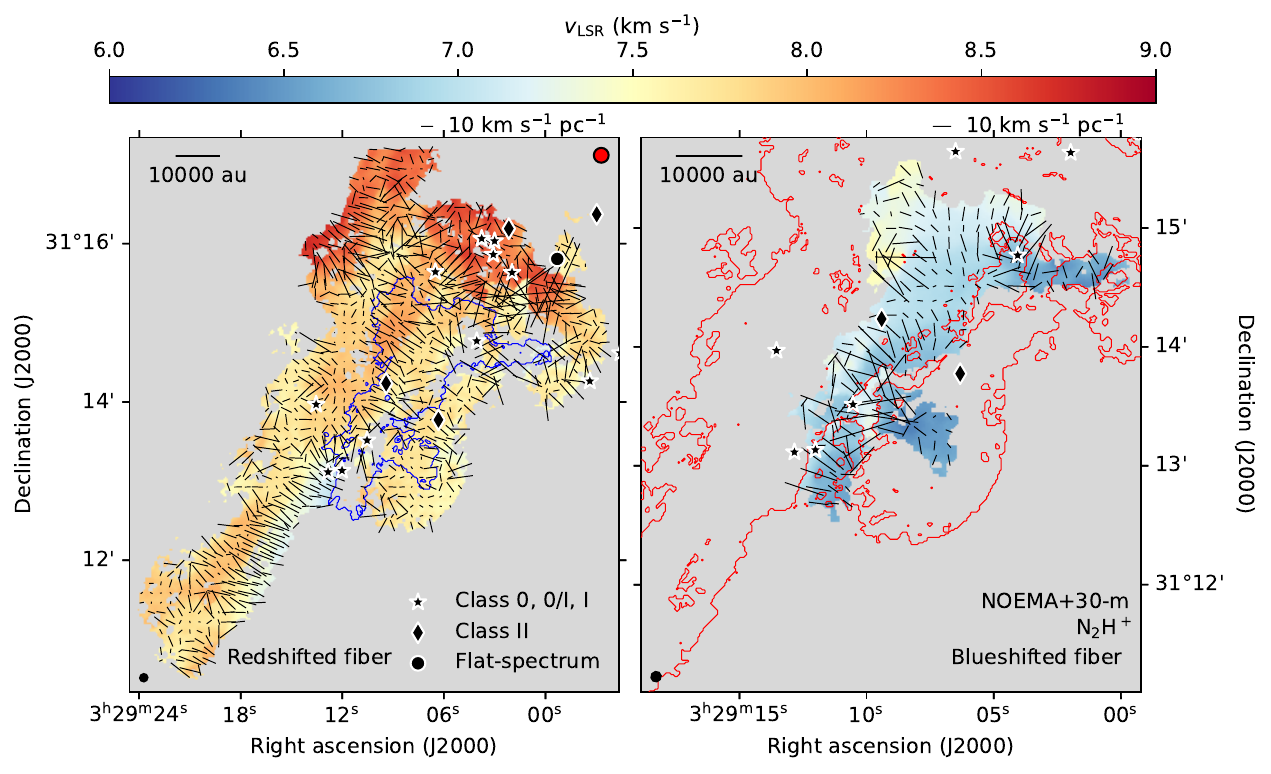}
	\includegraphics[width=0.94\textwidth]{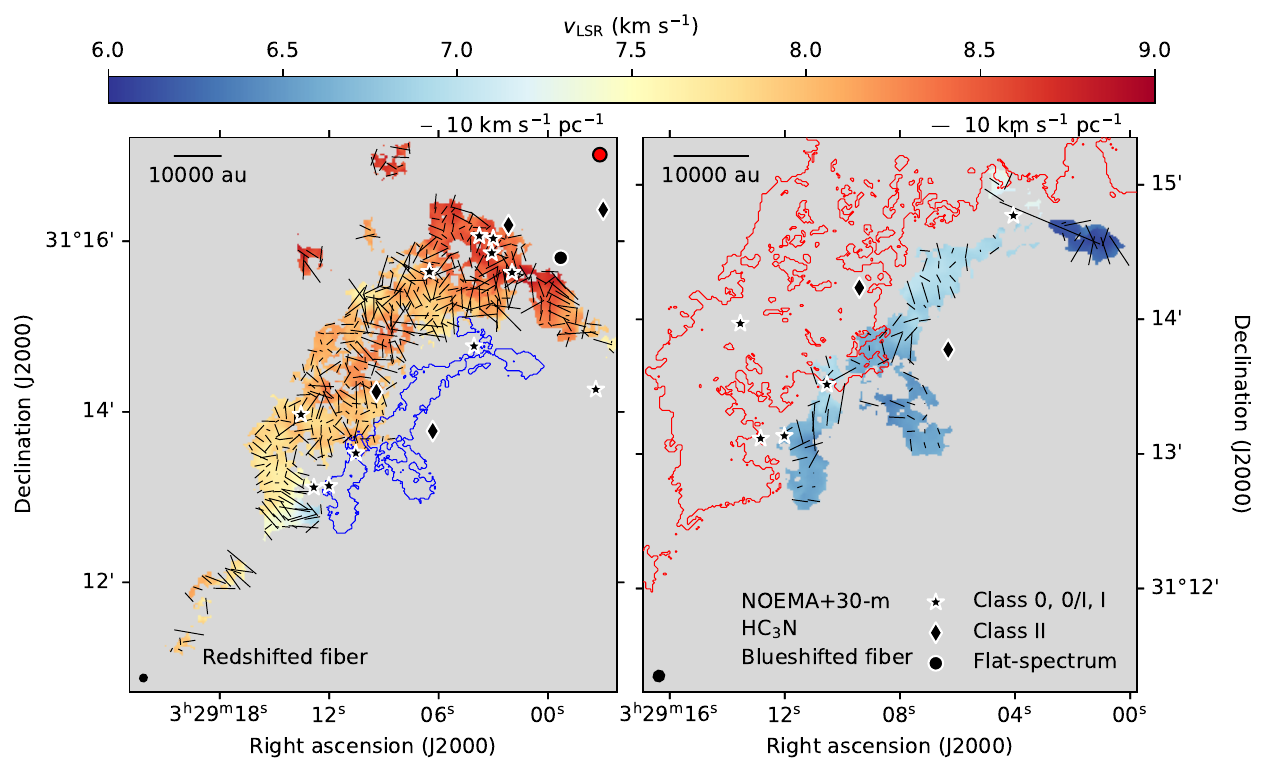}
	\caption{\refereeone{Resulting velocity groups for \ce{N2H^+} and \ce{HC3N} emission after the clustering process. The left plots show the \vlsr of the redshifted fiber, whereas the right plots show the same for the blueshifted fiber.} The blue contour marks the area occupied by the blueshifted fiber and the red contour, the redshifted fiber. The black symbols represent protostellar objects in the region as in Fig. \ref{fig:mom8maps}. The black ellipse at the bottom-left corner of both plots represents the beam size. \draftone{The black lines represent the velocity gradient directions measured at each position, with their size representing the gradient magnitude. The red circle at the top right corner of the left plot shows the size of the sampled area to calculate the gradients.} The panels do not have the same spatial scales. Top: \ce{N2H^+} emission \vlsr for each of the clusters. Bottom: \ce{HC3N} emission \vlsr after grouping the clusters according to their velocities with respect to the \ce{N2H^+} clusters.}
	\label{fig:HC3N-N2Hp-subfilaments-clusters}
\end{figure*}

\subsubsection{\refereeone{Velocity structure of  \ce{N2H^+} and \ce{HC3N}}}

We recovered the fiber structure observed in previous works with HDBSACAN (Sect. \ref{sec:clustering}) in \ce{N2H^+} emission. HDBSCAN recovers two clusters: we named cluster 0 the blueshifted  (blue) fiber and cluster 1, the redshifted (red) fiber as these resemble the fibers found in \cite{Chen2020b-velstructureNGC1333}, where one has a larger \vlsr than the other. \refereeone{The distribution of the clusters is more similar to the fibers found in \ce{NH3} by  \cite{Chen2020b-velstructureNGC1333} than for previous \ce{N2H^+} decomposition by \cite{Hacar2017NGC1333}. We discuss the differences between the structure characterization in our work with respect to \citep{Hacar2017NGC1333} in Appendix \ref{ap:fiber-comparison}.}  
Figure \ref{fig:HC3N-N2Hp-subfilaments-clusters} \refereeone{top} shows the \vlsr of each of the fibers. The redshifted fiber has an average $<\vlsr>=7.87$ \kms and the blueshifted fiber, $<\vlsr>=6.95$ \kms. The red cluster captures additional emission toward the northeast of the map, part of the Northeast (NE) filament \citep[as named by][]{Dhabal2019NGC1333NH3}, and toward the west, which is part of the extended emission not covered by our observations but observed in \ce{N2H^+} emission \citep{Hacar2017NGC1333}.

The clustering in \ce{HC3N}, \refereeone{unlike \ce{N2H^+}}, separates the emission into 7 clusters. There are no HDBSCAN parameters that give a similar clustering to \ce{N2H^+}. We grouped the \ce{HC3N} clusters into redshifted and blueshifted groups according to their velocity with respect to the average velocity of the \ce{N2H^+} clusters, so as to compare their velocities (Sect. \ref{sec:diff-vel}). Clusters in \ce{HC3N} which have an average velocity closest to the redshifted \ce{N2H^+} fiber are altogether called redshifted fiber, and clusters closer in velocity to the blue \ce{N2H^+} are grouped under the blueshifted fiber. The \vlsr of the resulting groups are shown in Fig. \ref{fig:HC3N-N2Hp-subfilaments-clusters}.

\subsubsection{Correlation of \ce{HC3N} emission with outflows\label{sec:hc3n-outflows}}

\begin{figure*}
    \centering
    \includegraphics[width=0.60\textwidth]{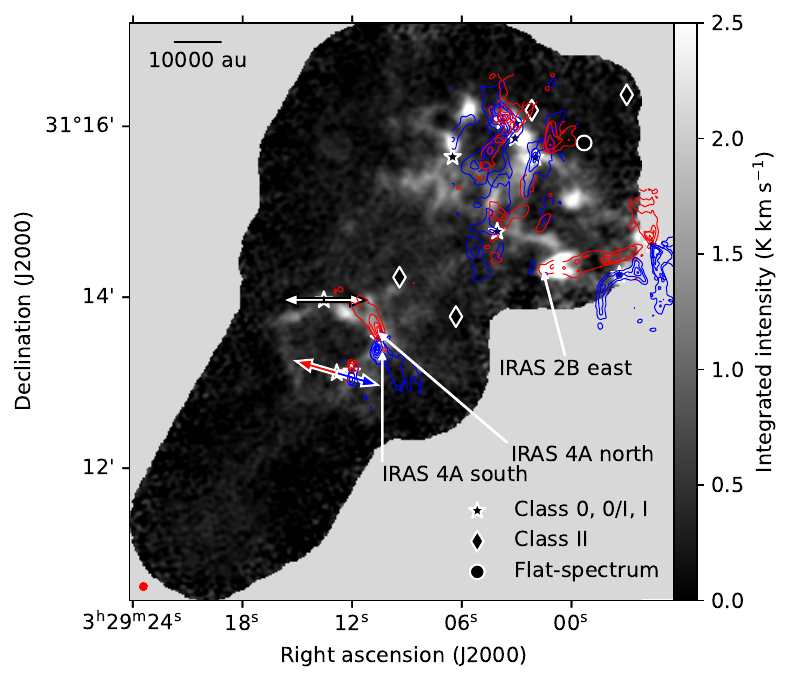}
    \includegraphics[width=0.38\textwidth]{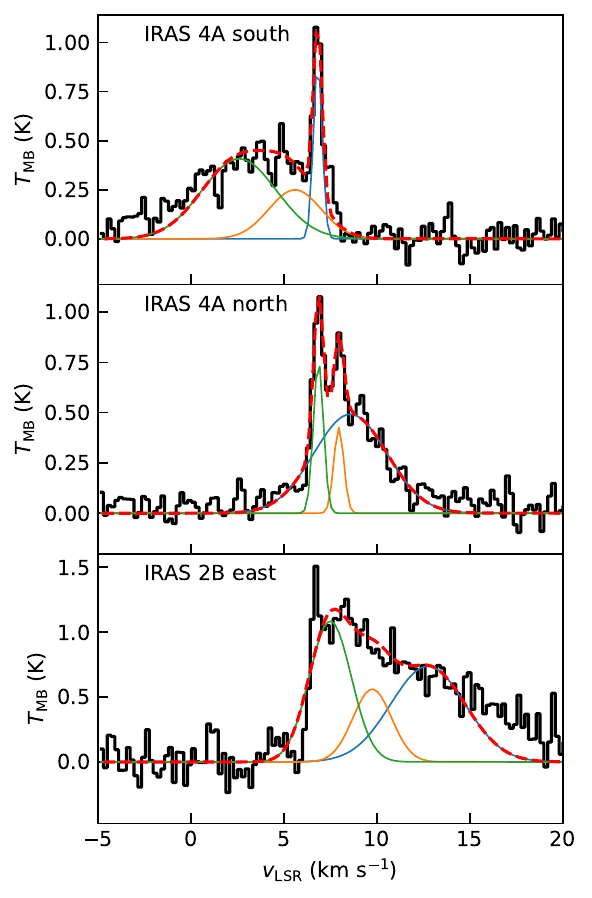}
    \caption{Correlation between the \ce{HC3N} emission and outflows. \draftone{Left: Integrated intensity map of \ce{HC3N} ($10-9$) between 5 and 10 \kms. The red symbols represent protostellar objects in the region as in Fig. \ref{fig:mom8maps}. The red and blue contours correspond to the \ce{^{12}CO} red and blue outflow lobes, respectively, obtained from the MASSES survey \citep{Stephens2019MASSES}.  The labels indicate where each of the spectra is taken from. Right:} \ce{HC3N} ($10-9$) spectra at the locations of the IRAS 4A outflow and at the bright emission to the east of IRAS 2B. For each location, we take the spectrum at an individual pixel. The blue, green and red curves represent the individual three Gaussians fitted at each position, and the red dashed curve represents the sum of all the Gaussians.}
    \label{fig:HC3N-outflows}
\end{figure*}

\draftone{Figure \ref{fig:HC3N-outflows} left shows the integrated intensity of \ce{HC3N} together with the \ce{^{12}CO} $J=2-1$ integrated emission maps toward the protostars, available from the MASSES survey  \citep{Stephens2019MASSES}.} We \draftone{show} the spectra at the locations where there are three Gaussians in the \ce{HC3N} cube (toward the north and south of IRAS 4A and toward the east of Per-emb 36) in Fig.~\ref{fig:HC3N-outflows} right. The spectra at these locations present extended wings: at the south of IRAS 4A, the wing is blueshifted with respect to the emission peak, and toward the north of IRAS 4A and the east of IRAS 2B, the wing is redshifted. All of these wings were fitted with one or two broad ($\sigma_{\mathrm{v}}> 2$ \kms) Gaussian components.  \draftone{The presence of these wings and their velocity (blueshifted or redshifted) coincide with the outflows shown in \ce{^{12}CO} $J=2-1$. }

\draftone{The wings in the spectra close to IRAS 4A are caused by the outflow driven by IRAS 4A, but} the bright \ce{HC3N} redshifted wing to the east of IRAS 2B coincides with the tip of the outflow lobe of IRAS 2A (also known as Per-emb 27). IRAS 2A is a binary system that has two outflows, one collimated in the east-west direction and a wider one in the north-south orientation \citep{Plunkett2013outflows}. The redshifted lobe of the collimated outflow is toward the east and shows intense  \ce{^{12}CO}  emission at the same location of \draftone{the brightest peak of the \ce{HC3N} integrated intensity map.} Part of this redshifted \ce{^{12}CO} emission is at the eastern edge of the IRAS 2B MASSES image and the other part is at the western edge of the Per-emb 15 MASSES map. As this emission does not coincide with other protostars or peaks in \ce{N2H^+} emission, we conclude this \ce{HC3N} enhancement is a bow shock due to the outflow from IRAS 2A impacting the cloud material. In the \ce{HC3N} channel maps, between 6.5 and \refereeone{8.8} \kms (Fig.~\ref{fig:HC3N-chanmap}), the images show a v shape at the position \refereeone{marked in Fig. \ref{fig:HC3N-outflows}  as IRAS 2B east}. We suggest this is due to the collision of the outflow lobe with cloud material, \draftone{which} produces a bow shock in the gas and enhances the presence of \ce{HC3N}.

Not all of the outflows cause wings in \ce{HC3N} emission. \ce{HC3N} is enhanced approximately along the IRAS 4C and IRAS 4B2 outflows (Fig. \ref{fig:HC3N-outflows}), but the directions of the enhancements are not over the position angles of the corresponding outflows. In the SVS 13 system, \ce{HC3N} emission is brighter at the \draftone{locations} of the protostars and extends all along the ridge that joins the protostars together, but does not present extended wings in the spectra. Moreover, comparing the \ce{^{12}CO} outflows from MASSES with the integrated \ce{HC3N} emission (without the outflow wings) in Fig.~\ref{fig:HC3N-outflows}, \ce{HC3N} seems to be brighter \textit{outside} of the regions covered by \ce{^{12}CO}, with the exception of the IRAS 2A and IRAS 4A outflows. Therefore, the outflows influence \ce{HC3N} emission differently depending on the location, but in general, \ce{HC3N} seems to be enhanced at the outflow cavities, while tracing extended material in general.

\subsubsection{Comparison between \ce{HC3N} and \ce{N2H^+} velocities\label{sec:diff-vel}}

We compared the extent and \vlsr of both molecules independently for each fiber. \ce{N2H^+} is used as a proxy for the dense gas, as in Perseus \ce{N2H^+} correlates well with locations of dense cores, where CO is depleted \citep{Johnstone2010nitrogendensegassperseus}. Previous works in this region show that \ce{N2H^+} follows the same structures as dust and \ce{NH3}, which means that \ce{N2H^+} traces of the physical structure of each fiber \citep{Hacar2017NGC1333,Dhabal2019NGC1333NH3}.  \ce{HC3N} is a known "early type" molecule together with other carbon-chain molecules (such as CCS and cyclic-\ce{C3H2}), meaning that in chemical models of molecular cloud collapse, these molecules appear at early times in comparison to, for example, nitrogen-bearing molecules \citep{Suzuki1992chem-carbon-nitrogen, Bergin-Tafalla2007}. 
We used the channel maps together with the spectral decomposition made in Sect. \ref{sec:methods} to describe the gas structure.

In general, \ce{HC3N} structure consists of \draftone{several small peaks in emission connected via less bright, extended emission}. 
\ce{HC3N} does not cover the full extent of the \draftone{NGC 1333 SE} filament nor each of the \ce{N2H^+} fibers: the area covered by \ce{HC3N} emission is smaller than the one covered by \ce{N2H^+} emission, as seen in Fig. \ref{fig:res-ncomp}, and is particularly scattered toward the filament tail in the south. \ce{HC3N} is detected inside the area defined for both fibers with \ce{N2H^+} emission: in the redshifted fiber, \ce{HC3N} is roughly detected along the center, whereas in the blueshifted fiber \ce{HC3N} is detected only toward the west side of the fiber. 

\begin{figure*}
    \centering
    \includegraphics[width=0.64\textwidth]{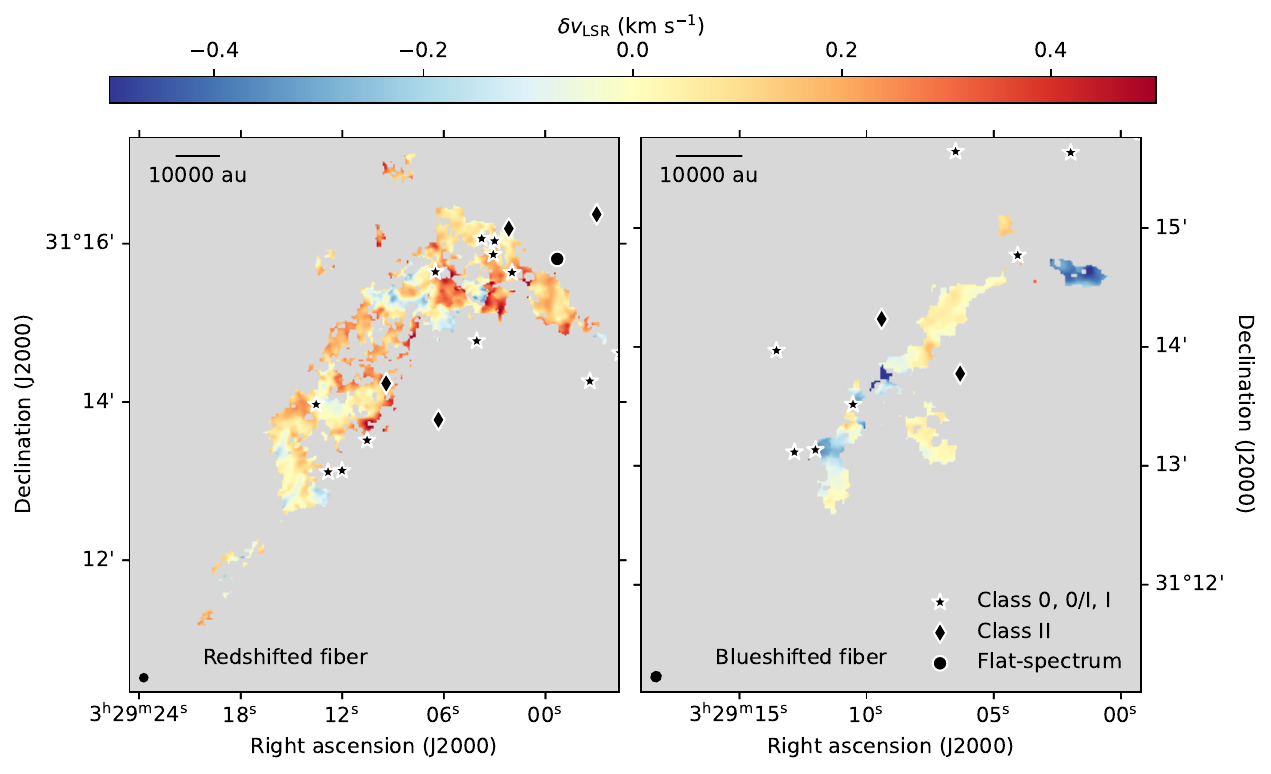}
    \includegraphics[width=0.34\textwidth]{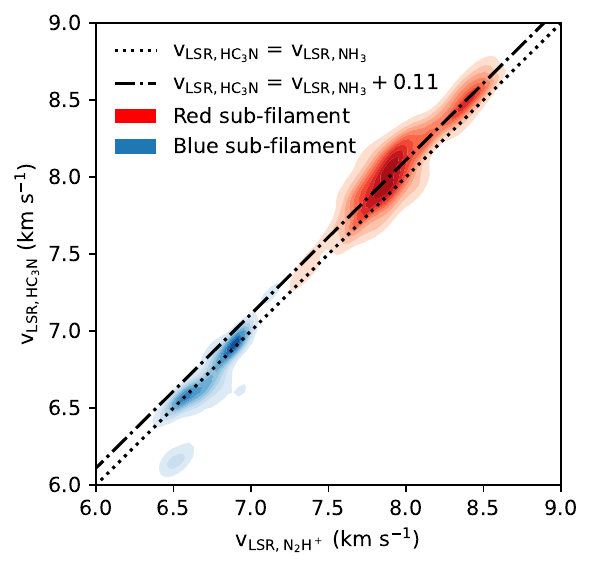}
    \caption{Difference between the velocity components of each fiber in NGC 1333 SE, as recognized after the clustering. Left: Resulting difference after subtracting the \ce{N2H^+} \vlsr from the \ce{HC3N} \vlsr of the clusters belonging to the redshifted fiber. Middle: Resulting difference after subtracting the \ce{N2H^+} \vlsr from the \ce{HC3N} \vlsr of the clusters belonging to the blueshifted fiber. Note that the blueshifted fiber map is zoomed into the region covered by the emission. Right: two-dimensional KDE of the \ce{HC3N} \vlsr versus \ce{N2H^+} \vlsr. The dotted line represents the location where the velocities are equal, and the dash-dotted line, where the \ce{HC3N} \vlsr is larger by 0.11 \kms (the median difference for the redshifted fiber). }
    \label{fig:diff-bluered}
\end{figure*}

We compared the velocities traced by both molecules in each fiber. 
\draftone{Figure \ref{fig:diff-bluered} left shows the velocity difference between  \ce{HC3N} and \ce{N2H^+} $\delta \vlsr = \rm v_{\mathrm{LSR, HC}_3\mathrm{N}} - \rm v_{\mathrm{LSR, N}_2\mathrm{H}^+}$.} We reprojected the \ce{HC3N} images of each velocity component to the pixel grid of the \ce{N2H^+} cube, because the pixel size and beam is larger. Most of the redshifted fiber has \ce{HC3N} velocities that are larger \draftone{than the velocity of} \ce{N2H^+} gas. The exception to this is a band that runs east to west below the SVS 13 area. The blueshifted fiber shows considerable differences between the molecules' velocities close to IRAS 4A and 4B, but in the rest of the gas, there is no apparent difference.

We evaluated if the values of $\delta \vlsr$ are significant using the Kernel Density Estimate (KDE) of the differences between the \vlsr of both molecules. This KDE is in Fig.~\ref{fig:diff-bluered} right. The median difference is 0.11 \kms for the redshifted fiber and 0.01 \kms for the blueshifted fiber. $\delta \vlsr$ for the red fiber is larger than the sum of the uncertainties in both $\rm v_{\mathrm{LSR, HC}_3\mathrm{N}}$ and $\rm v_{\mathrm{LSR, N}_2\mathrm{H}^+}$ obtained from the Gaussian fitting (0.08 \kms), \draftone{and therefore show that the gas in both molecules presents different velocities.} This is not the case for the blue fiber. 

\subsubsection{Velocity gradients within the fibers\label{sec:vel-gradients}}

We calculated the gradients present in the \ce{N2H^+} and \ce{HC3N} velocity fields. The local velocity gradients describe the fluctuations of gas motion within the fibers, and are of assistance to identify infall toward individual \refereeone{YSOs} or binaries. We used a similar approach as used in \cite{Chen2022B5} and \cite{Valdivia-Mena2023B5}. We calculated the local velocity gradient by fitting a plane centered at one pixel, with a width of 2 beams, so that we capture gradients in uncorrelated pixels. The plane is defined as:
\begin{equation}
    \vlsr = Ax + By + v_c,
\end{equation}
where $A$ is the slope in the x (right ascension) direction and $B$ in the y (declination) direction. \draftone{We used the velocity gradient implementation within the \texttt{velocity\_tools} package\footnote{https://github.com/jpinedaf/velocity\_tools}.}

The resulting gradient \draftone{orientations} and magnitudes are plotted over the \vlsr of each fiber in Fig.~\ref{fig:HC3N-N2Hp-subfilaments-clusters} for \ce{HC3N} and \ce{N2H^+}. In both fibers, the velocity gradient field for \ce{HC3N} is more randomly aligned than for \ce{N2H^+}. This indicates that the gas traced by \ce{HC3N} presents larger changes at local scales, \draftone{in part} due to the effect of outflows in the region. 

Both  \ce{N2H^+} and \ce{HC3N} show large (about 20 \kms pc$^{-1}$) gradients within the fibers, almost completely perpendicular to the filament, in regions where there are no protostars. \refereeone{In particular, the velocity gradients toward the south of the redshifted fiber in both molecules seem well ordered.}  
Between IRAS 4 and SVS 13, the velocity gradients between both molecules are different. The \ce{N2H^+} gradient vectors are \refereeone{mostly} perpendicular to the length of the filament in the blueshifted fiber, but with a lower magnitude ($\lesssim10$ \kms pc$^{-1}$). In the redshifted fiber, there is a region with a sudden change in velocity that produces local, perpendicular gradients of about 20 \kms pc$^{-1}$, but in the rest of the region the local gradients are smaller than 5 \kms pc$^{-1}$ and do not show clear patterns. On the other hand, \ce{HC3N} emission presents strong variations in the redshifted fiber, up to 30 \kms pc$^{-1}$, but apparently randomly oriented, and roughly perpendicular with magnitudes between 10 and 20 \kms pc$^{-1}$ in the blueshifted fiber.

\subsection{Streamer candidates\label{sec:streamers}}

We analyzed the molecular tracers surrounding the \refereeone{YSOs} in search for signatures of streamers. The observations in this work are designed to follow the flow from the larger scales of the filaments to the smaller, protostellar scales. We list the general properties of the \refereeone{YSOs} within our field of view in Table \ref{tab:protostars}.

The search for streamers in the region was done using the following signatures: we first looked for velocity gradients in our \ce{HC3N} and \ce{N2H^+} maps where the difference between the \vlsr of \draftone{our observations} and the protostar increases as the distance between the gas and the protostar decreases. This is a common characteristic of streamers observed towards protostars \refereeone{and pre-main sequence stars} \citep[e.g.][]{Thieme2022Lupus3streamers,Valdivia-Mena2022per50, Valdivia-Mena2023B5, Hsieh2023svs13astreamer}. We analyzed both the maps of the velocity gradients within clusters (Fig. \ref{fig:HC3N-N2Hp-subfilaments-clusters}) as well as the central velocities fitted to regions that were not clustered in Sect. \ref{sec:clustering}. \draftone{For embedded protostars (Class 0, 0/I and I), we used the \vlsr reported in the MASSES survey \citep{Stephens2019MASSES} as protostellar velocity, obtained using a Gaussian fit to SMA \ce{C^{18}O} (2-1) emission observations within an area of 1.2\arcsec. If the velocity was not available in MASSES, we used the \vlsr reported in single-dish observations of DCN \citep{Imai2018deuterium}. For Class II sources, we used reported \vlsr values from APOGEE spectra \citep{Foster2015virialcores-prems, Kounkel2019closecompanions}.} When the velocity of the protostars not available from \draftone{the} previous observations, we \draftone{adopted the velocity of \ce{N2H^+} at the source location.} Then, we determined if the velocity gradient comes from a preferential direction and is not a radial velocity gradient centered at the protostar. Finally, we observed if there is any elongated structure, such as bright lanes or peaks not centered at the protostar's location, in $T_{\mathrm{MB}}$.  \draftone{We defined the region that makes up the potential streamer manually, while including emission within the S/N $=10$ contours.}

We \draftone{constructed} sub-cubes of \ce{HC3N} and \ce{N2H^+} emission and images of the Gaussian component properties (Sect.~\ref{sec:bayesian}) and clusters (Sect.~\ref{sec:clustering}), centered at each protostar and including all pixels within a 10, 000 au radius. We chose 10, 000 au based on the longest confirmed streamer found to date \citep{Pineda2020Per2}, which is characterized using \ce{HC3N} emission. We \draftone{excluded from} our analysis the broad components correlated with outflow activity seen in Sect.\ref{sec:hc3n-outflows}. We expected \ce{HC3N} to reveal streamer motion, as streamers have been mostly observed in carbon-bearing molecules \citep{Pineda2020Per2}, and \ce{N2H^+} can give information about the surrounding envelope, \draftone{but we also checked the \ce{N2H^+} velocity gradients}. The beam of our \ce{HC3N} data is 4.9\arcsec, which corresponds to a length of almost 1500 au. Based on this, we are able to determine the presence of streamers (or asymmetric infall) using \ce{HC3N} that are at least $\sim 1500$ au in projected length. We note that streamers can be longer in reality, but we are limited by their projected length in the plane of the sky unless we are able to model their infall kinematics, which was evaluated in each individual case. 
Therefore, our \ce{HC3N} data has the potential to reveal asymmetric infall of sizes typical of previously discovered streamers toward Class 0 and I protostars \citep{Thieme2022Lupus3streamers, Valdivia-Mena2022per50, Valdivia-Mena2023B5}. There is no detected \ce{HC3N} emission around ASR 53, EDJ2009-183 (ASR 106) and EDJ2009-173 (ASR 118), and Per-emb 27 (IRAS 2A) is out of the imaged area, so these sources are not analyzed. 

\begin{figure*}
    \centering
    \includegraphics[width=0.44\textwidth]{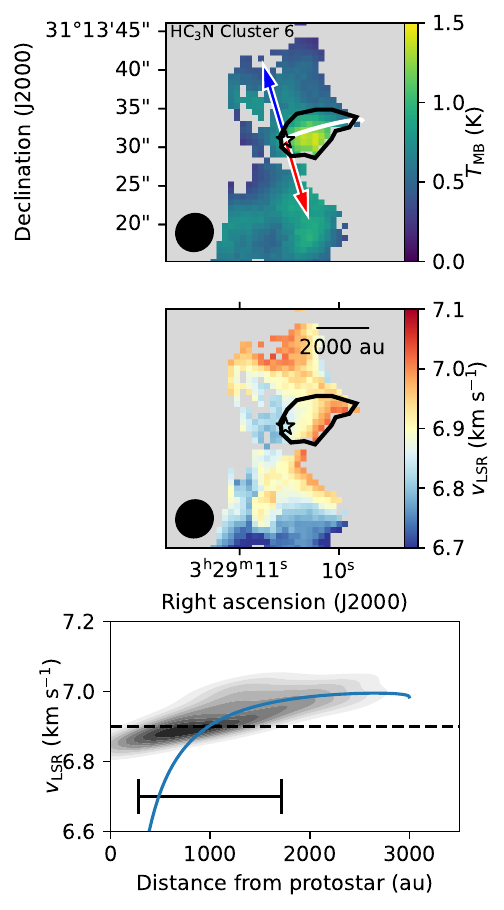}
    \includegraphics[width=0.45\textwidth]{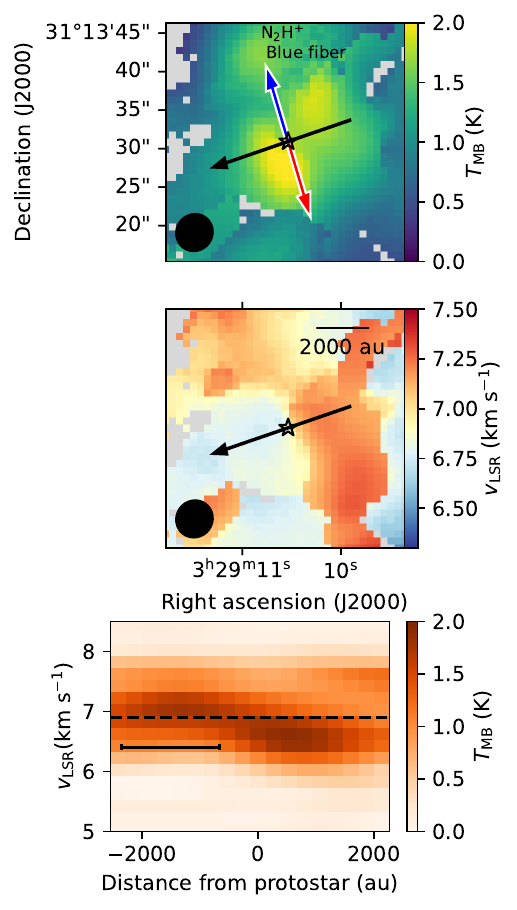}
    \caption{Zoom-in plots of \ce{HC3N} \refereeone{(left) and \ce{N2H^+} (right)} emission for IRAS 4A, together with the best free-fall trajectory found for the $T_{\mathrm{MB}}$ and \vlsr maps. Top: Amplitude $T_{\mathrm{MB}}$ of the Gaussian components corresponding to \ce{HC3N} cluster no. 6 \refereeone{and to \ce{N2H^+} blue cluster}. The black polygon marks the region selected as a potential streamer. The white curve marks the potential streamer's trajectory. The blue and red arrows indicate the direction of the outflow lobes for IRAS 4A and 4B, \refereeone{and the black arrow over the \ce{N2H^+} map shows the orientation of the position-velocity diagram at the bottom}. Middle: central velocities \vlsr of the Gaussian component. A scalebar representing 2000 au is at the top right of the image. Bottom \refereeone{left}: KDE of the \vlsr within the selected region. The black line at the \refereeone{bottom} of the plot represents a length \refereeone{of} one beam. The blue curve marks the velocity versus the distance for the \refereeone{found free-fall solution}. \draftone{The black dashed line represents the protostar's \vlsr}. \refereeone{Bottom right: \ce{N2H^+} position-velocity diagram along the path indicated in the top panel. The dashed horizontal line marks the \vlsr of the source \citep[6.9 \kms,][]{Stephens2019MASSES}. \refereeone{The black scalebar represents a length equivalent to one beam.}}}
    \label{fig:zoom-iras4a-hc3n-n2hp}
\end{figure*}

We found streamer candidates toward 7 \refereeone{out of the 16} \refereeone{YSOs} from Table \ref{tab:protostars}. In the following, we present the analysis of each individual protostar that has a streamer candidate. We present IRAS 4A first as it is the only one where there is enough resolution and \refereeone{available} information about its stellar properties to confirm its infall motion using a free-fall \draftone{model}. \draftone{Figure \ref{fig:zoom-iras4a-hc3n-n2hp} shows the peak brightness \refereeone{and velocity of \ce{HC3N} and  \ce{N2H^+}, together with} a streamer trajectory solution zoomed at the selected 10, 000 au scale. Figure \ref{fig:zoom-hc3n-all} shows the same \refereeone{\ce{HC3N}} properties (except for the free-fall model) for the rest of the streamer candidates, \refereeone{and Fig. \ref{fig:zoom-n2hp-all} shows their \ce{N2H^+} observed properties.}} We found no evident streamer candidates in IRAS 4B2, IRAS 4C (Per-emb 14), SVS 13B, SVS 13C and ASR 3.

\begin{figure*}
    \centering
    \includegraphics[width=0.98\textwidth]{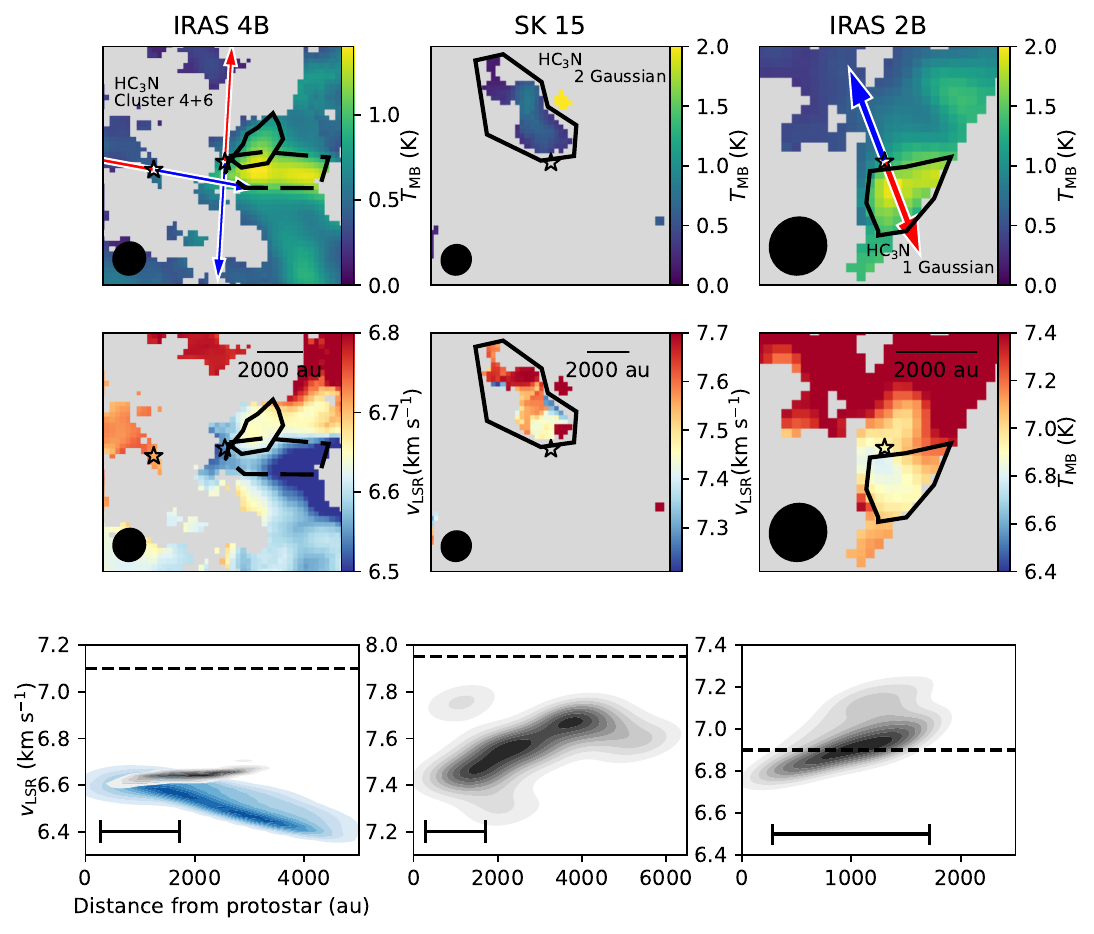}
    \caption{Zoom-in plots of \ce{HC3N} emission for IRAS 4B (left), SK 15 (center) and IRAS 2B (right), with the same panels as shown in Fig. \ref{fig:zoom-iras4a-hc3n-n2hp} \refereeone{left}. The Gaussian component for each protostar is labeled accordingly. The black polygon marks the region selected as a potential streamer. The black ellipse at the bottom left corner represents the beam. For IRAS 4B, the \refereeone{dashed black} polygon represents the region analyzed in relation to IRAS 4B2. Top: Amplitude $T_{\mathrm{MB}}$ of the Gaussian component plotted. The blue and red arrows indicate the direction of the blueshifted and redshifted outflow lobes, respectively, for known outflows in the plotted area.  Middle: Central velocity \vlsr of the Gaussian component selected. The scalebar represents a length of 2000 au. Bottom: KDE of the \vlsr within the selected region. The black density histograms represent the KDE of the velocities within the black polygons. The dashed lines mark the \vlsr of each protostar. The black scalebar represents a length equivalent to one beam. For IRAS 4B, the blue KDE represents the velocities within the \refereeone{dashed} polygon. }
    \label{fig:zoom-hc3n-all}
\end{figure*}

\begin{figure*}
    \ContinuedFloat
    \captionsetup{list=off,format=cont}
    \centering
    \includegraphics[width=0.98\textwidth]{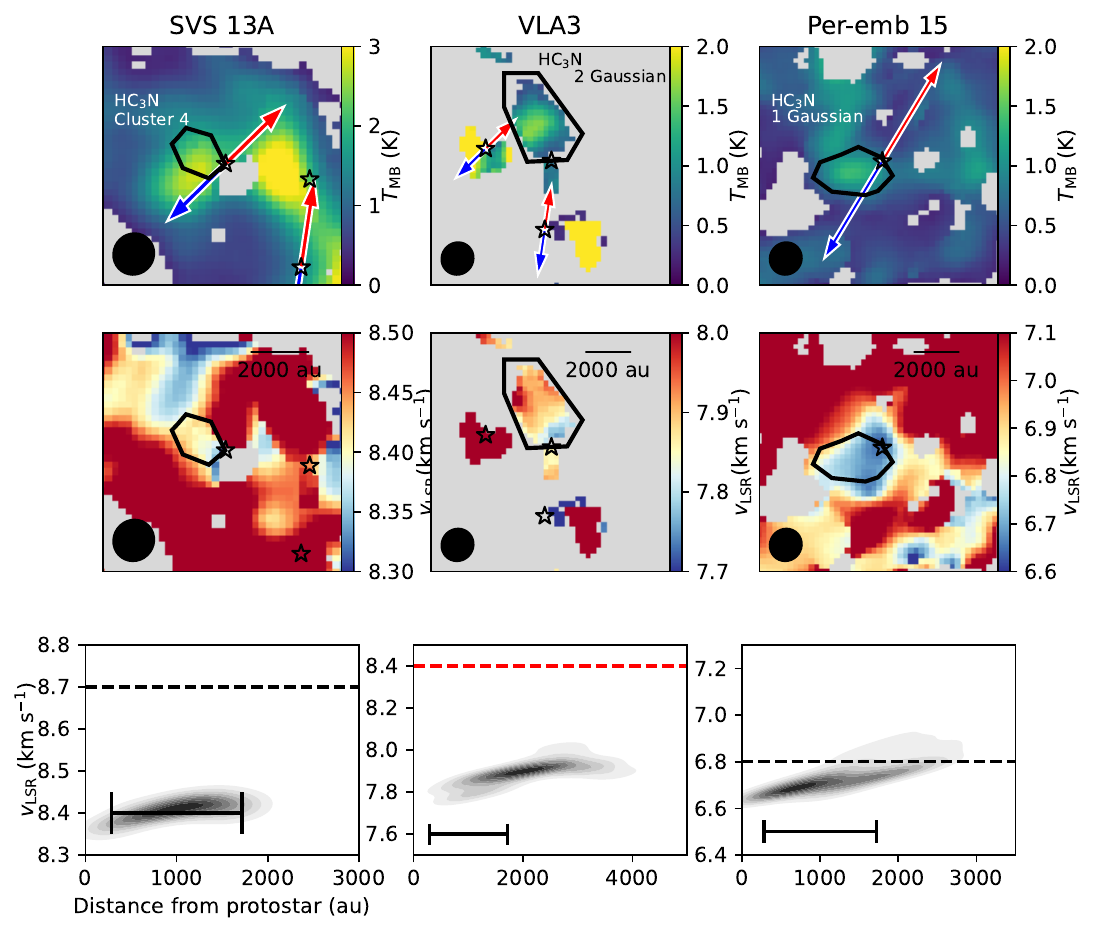}
    \caption{Zoom-in plots of \ce{HC3N} emission for SVS 13A (left), VLA 3 (center) and Per-emb 15 (right), with the same panels as shown in Fig. \ref{fig:zoom-iras4a-hc3n-n2hp} \refereeone{left}. For VLA 3, the red dashed line marks the \vlsr of the fiber component toward the position of the protostar.}
\end{figure*}

\subsubsection{IRAS 4A\label{sec:iras4a}}

IRAS 4A is a Class 0 protostar that drives a strong outflow, described in several previous works \citep[e.g.,][]{Plunkett2013outflows}. We also observed the effect of this outflow in the \ce{HC3N} emission (Sect. \ref{sec:hc3n-outflows}). We examined the different cutout images for the surroundings of IRAS 4A and found a region with a velocity gradient consistent with infall motion. There is an increase in $T_{\mathrm{MB}}$ toward the west of the protostar \draftone{in the \ce{HC3N} cluster no. 6, part of the blueshifted fiber}. We show the $T_{\mathrm{MB}}$ of this cluster in Fig.~\ref{fig:zoom-iras4a-hc3n-n2hp} top \refereeone{left}. This region \draftone{has} a velocity gradient from west to east, which we show in Fig.~\ref{fig:zoom-iras4a-hc3n-n2hp} bottom \refereeone{left}: at approximately 2500 au from IRAS 4A, \ce{HC3N} gas has a $\vlsr = 7.0$ \kms, whereas closer to the protostar it reaches $\vlsr = 6.85$ \kms. The uncertainty in the fitted velocity is 0.03 \kms on average for this region, so the difference between both extremes is significant. Therefore, we defined this region as a streamer candidate in \ce{HC3N} emission.

We were able to confirm that this velocity profile is consistent with free-fall infalling gas toward the protostar. Using the methods detailed in \cite{Pineda2020Per2} and \cite{Valdivia-Mena2022per50}, we modeled the velocity gradient in the image plane and the velocity plane, assuming a free-fall, ballistic trajectory from \cite{Mendoza2009}. \draftone{We retrieved the} disk inclination from the VANDAM survey results \citep{Segura-Cox2018AVANDAM} \draftone{and the position angle from the MASSES survey \citep{Lee2016MASSES}}. We only used the mass of the envelope \draftone{determined} in \cite{Jorgensen2007SMAPerseusenv}, $M_{\mathrm{env} }= 2.3$ \Msun \draftone{for the total mass of the system}, which is a good enough approximation to model the free-fall for a Class 0 protostar, as most of the mass is in the envelope. Table \ref{tab:paramsstream} shows the set of parameters that can approximately reproduce the KDE of the observed \ce{HC3N} velocities within a region that surrounds the $T_{\mathrm{MB}}$ peak area. The resulting trajectory \refereeone{from these parameters} is plotted in Fig.~\ref{fig:zoom-iras4a-hc3n-n2hp} \refereeone{left}. We note that this is a rudimentary adjustment of these velocities, and that the region we believe the streamer to cover has an equivalent size of two beams. \refereeone{At scales smaller than $\sim1000$ au from the protostar, the best free-fall solution diverges from the observed \vlsr distribution as there are possibly other gas components within the beam (for instance, protostellar envelope and disk rotation). Our current resolution limits our interpretation to scales larger than about 1500 au.} Higher resolution observations of \ce{HC3N} line emission \draftone{will allow for a better trajectory modelling}.

\begin{table}[ht]
\centering
\caption{\label{tab:paramsstream}Parameters of the trajectory that  best reproduce the \ce{HC3N} observations around IRAS 4A.}

\begin{tabular}{lll}
\hline\hline
Parameter  & Unit   & Value     \\ \hline
$\vartheta_0$ & deg   & 92 \\
$\varphi_0$   & deg    & -5  \\
$r_0$      & au     & 3000 \\
${\rm v}_{r,0}$  & \kms   & 0  \\
$\Omega_0$ & s$^{-1}$ & $2.3 \times 10^{-13}$ \\
$i^*$        & deg   & -35    \\
P.A.$^{**}$       & deg  & 19  \\ \hline
\end{tabular}
\tablefoot{\tablefoottext{*}{Angle obtained from the disk inclination \citep{Segura-Cox2018AVANDAM}.}
\tablefoottext{**}{Angle obtained from the outflow direction \citep{Lee2016MASSES}.
}}
\end{table}

The \ce{N2H^+} velocity profile is significantly different than what \ce{HC3N} shows in this region. IRAS 4A is located at $\approx 4$\arcsec from a local \ce{N2H^+} peak, as seen in Fig.~\ref{fig:zoom-iras4a-hc3n-n2hp} top \refereeone{right}. There is a velocity gradient from redshifted to blueshifted velocity with respect to the protostar's \vlsr (6.9 \kms) going from west to east (Fig.~\ref{fig:zoom-iras4a-hc3n-n2hp} middle \refereeone{right}). The eastern side of this gradient is blueshifted with respect to the protostar's \vlsr and is brighter than its redshifted counterpart. We made a position-velocity (PV) cut perpendicular to the outflow direction, shown in Fig.~\ref{fig:zoom-iras4a-hc3n-n2hp} bottom \refereeone{right}. The shape of the \ce{N2H^+} PV diagram is suggestive of rotation within the envelope surrounding the protostar. IRAS 4A is expected to be within an envelope as it is classified as a Class 0 source. In previous works, gas surrounding IRAS 4A was found to be infalling \citep{Belloche2006compressedinfallIRAS4A}. This profile then could be associated with a rotating-infalling envelope.

\subsubsection{IRAS 4B\label{sec:iras4b}}

IRAS 4B is a Class 0 protostar that is to the southeast of IRAS 4A. Together with IRAS 4B2 they form a binary system. There is a bright \ce{HC3N} emission perpendicular to the outflow that dominates the emission around IRAS 4B, marked \refereeone{with a dashed black polygon} in Fig.~\ref{fig:zoom-hc3n-all} left. The direction of the lane coincides with the direction of the IRAS 4B2 outflow (Fig.~\ref{fig:zoom-hc3n-all} left) and the velocity of this lane increases with increasing distance, so this bright lane is probably not asymmetric infall. 
However, at a $45\deg$ angle from this lane, there is another \draftone{signature of} extended \ce{HC3N} emission, fitted with one Gaussian and belonging to cluster no. 6, and therefore part of the blueshifted fiber seen in Sect. \ref{sec:clustering}. This smaller region is marked with a black polygon in Fig.~\ref{fig:zoom-hc3n-all} top left. This bright emission has a velocity gradient that is consistent with infall toward IRAS 4B, based on the KDE of the observed velocities in Fig.~ \ref{fig:zoom-hc3n-all} bottom left (black KDE). The difference between the \vlsr of the protostar and the \vlsr of the candidate \draftone{at the beginning of the streamer (at about 3000 au, Fig. \ref{fig:zoom-hc3n-all})} is about 0.4 \kms, which indicates that, to model this with a streamline model, we \draftone{may require} an initial radial velocity ${\rm v}_{r,0}\neq 0$. Nevertheless, the central velocity of this protostar has been reported between 6.8 and 7.1 \citep{Imai2018deuterium, Stephens2019MASSES}. We chose the value from \citep{Stephens2019MASSES} as it is the \draftone{one obtained with the best resolution}.

At the position of IRAS 4B, there is a local \ce{N2H^+} brightness peak. By doing a PV cut on the direction perpendicular to the outflow direction (Fig. \ref{fig:zoom-n2hp-all} left), we observed that this peak is slightly blueshfted with respect to the protostellar \vlsr. There is no sign of a rotating or infalling envelope in the \ce{N2H^+} surrounding either IRAS 4B or 4B2: both regions are dominated by emission \draftone{at} fiber \draftone{scales}.

\subsubsection{SK 15\label{sec:sk15}}

SK 15 \citep[from the nomenclature of][]{Sandell-Knee2001protostars} is a Class I protostar toward the southeast of SVS13A. Its \vlsr is estimated between 8 \kms \citep{Imai2018deuterium} and 8.1 \kms \citep{Higuchi2018chemchains}. To the best of our knowledge, it does not have an observed outflow of its own, but lies close to the outflow driven by SVS 13A (northern part of the map in Fig.~\ref{fig:HC3N-outflows}). 
We found a streamer \draftone{candidate} in \ce{HC3N} emission in the two-Gaussian fit around this protostar (Fig. \ref{fig:zoom-hc3n-all} center). One of the \ce{HC3N} Gaussian components is \draftone{part of cluster no. 4, within the redshifted fiber} (Fig. \ref{fig:cluster-results}), whereas the other is \draftone{recognized} as noise by HDBSCAN. This extra component is the weakest of the two in $T_{\mathrm{MB}}$ (0.5 K versus 2.2 K). Looking at the component that is not part of the larger filament, the tail-shaped extension has a velocity gradient that gets more blueshifted with respect to SK 15's \vlsr as gas gets closer to SK 15 (Fig.~\ref{fig:zoom-hc3n-all} center). 

SK 15 is located within 7\arcsec of a \ce{N2H^+} $T_{\mathrm{MB}}$ peak. This peak is located at the position of the streamer candidate (Fig. \ref{fig:zoom-n2hp-all} middle). In this peak, there are two Gaussian components in the \ce{N2H^+} spectral decomposition (Sect.~\ref{sec:bayesian}). The component responsible for the $T_{\mathrm{MB}}$ peak has a \vlsr of approximately 8.1 \kms and belongs to the redshifted fiber. The other component, not clustered by HDBSCAN, has similar \vlsr to the \ce{HC3N} emission that we consider as streamer candidate, but it does not show a velocity gradient. 

\subsubsection{IRAS 2B\label{sec:iras2b}}

IRAS 2B \citep[also known as Per-emb-36,][]{Enoch2009embeddedprotostars} is a Class I protostar located at the western edge of the ProPStar \draftone{map}. There is denser gas beyond the edge of the map, as shown in previous works covering this region in, for instance, \ce{NH3} \citep{Friesen2017GAS, Dhabal2019NGC1333NH3}. Towards this edge, we found a bright extension of \ce{HC3N} gas that has a gradient from redshifted to blueshifted \vlsr with respect to IRAS 2B's $\vlsr = 6.9$ \kms (Fig.~\ref{fig:zoom-hc3n-all} right). This region is about the size of the \ce{HC3N} beam. We consider this as a streamer candidate as it shows a \draftone{velocity} gradient almost perpendicular to the outflow direction, although the complete gradient is contained within a beam. The brightness distribution suggests the streamer candidate could continue beyond the extend of the \ce{HC3N} map.

The \ce{N2H^+} map also shows that the dense gas emission continues to the right of IRAS 2B. Previous \ce{N2H^+} observations from \cite{Dhabal2019NGC1333NH3} show that there is considerable emission outside our coverage, toward the west of IRAS 2B, so we only get a glimpse of the kinematics toward the east of IRAS 2B. Figure \ref{fig:res-ncomp} shows that toward the northwest and southeast of the protostar, \ce{N2H^+} emission can be fit with two Gaussians. We made a PV cut perpendicular to the outflow direction, shown in Fig.~\ref{fig:zoom-n2hp-all} right. Within 4000 au from the protostar at each side, we see no clear indication of a rotating envelope, so \draftone{we conclude that the} \ce{N2H^+} emission is dominated by the \draftone{fiber kinematics}.

\subsubsection{SVS 13A\label{sec:svs13a}}

SVS 13A, also known as Per-emb-44 \citep[using the nomenclature of][]{Enoch2009embeddedprotostars}, is a Class 0/I close binary system \citep[separated by approximately 70 au,][]{Anglada2000SVS13Abinary}. It drives powerful outflows and jets that dominate the CO emission of this region \citep{Plunkett2013outflows, Stephens2019MASSES}. 
\draftone{There is a detected streamer toward this source seen in DCN emission by \cite{Hsieh2023svs13astreamer}.}
The cut \ce{HC3N} images toward SVS 13A show a beam-sized region consistent with the velocity gradient shown by \cite{Hsieh2023svs13astreamer}. The \ce{HC3N} cluster no. 4 shows a sudden drop in \vlsr \draftone{at the position} of SVS 13A (Fig.~\ref{fig:zoom-hc3n-all} (cont.) left). The fiber's \vlsr around SVS 13A is between 8.4 and 8.5 \kms, and in the region where the streamer was found using DCN the velocity drops down to approximately 8.3 \kms. This change is similar to the velocity gradient observed in the SVS 13A streamer \citep[][in their Fig. 6]{Hsieh2023svs13astreamer}. However, the \ce{HC3N} brightness distribution is difficult to interpret as this particular streamer. As seen in Fig.~\ref{fig:zoom-hc3n-all} (cont.) top left, 
there is a local peak in $T_{\mathrm{MB}}$ located at 4.5\arcsec east from the SVS 13A. This peak might blend with the less bright \draftone{emission} from the streamer. It is also possible that the streamer is small, about the size of our beam ($\sim 1500$ au), and at this resolution, any \ce{HC3N} emission coming from the streamer is blended with the larger, fiber-dominated emission. Nevertheless, we were able to detect the small velocity gradient even in the blended image, which means that it is possible that \ce{HC3N} emission traces the streamer toward SVS 13A. A higher spatial resolution image should be able to \draftone{separate} the emission corresponding to the fiber and the streamer.

SVS 13A lacks an \ce{N2H^+} peak. The brightest peak near SVS 13A is located on SVS 13B to the southwest. We plot a PV diagram centered on SVS 13A and in the direction perpendicular to the outflow in Fig.  \ref{fig:zoom-n2hp-all} (cont.) left. \cite{Chen2009SVS13Bjointcore} show that SVS 13B and VLA 3 are embedded in the same core, which shows signs of rotation. Therefore, to the west, the \ce{N2H^+} velocity field is dominated by the SVS 13B and VLA 3 joint core. To the east, the velocity field is dominated by the extended \ce{N2H^+} component tracing the redshifted fiber.  

\subsubsection{VLA 3\label{sec:vla3}}

VLA 3 is a protostellar source located 10\arcsec toward the west of SVS 13A. It does not have an outflow direction on record, nor information about a potential disk. However, it lies on the \draftone{path} of the SVS 13B outflow. As mentioned in Sect. \ref{sec:svs13a}, VLA 3 and SVS 13B are embedded in a common core, suggesting these two sources are forming a bound binary system \citep{Chen2009SVS13Bjointcore}. As with the case of SK 15, one of the \ce{HC3N} Gaussian components located near VLA 3 is part of the redshifted fiber, within the HDBSCAN cluster no. 4. The other \ce{HC3N} Gaussian component is categorized as noise by the clustering algorithm. Looking at this remaining component, there is a velocity gradient that points toward VLA 3. We show the $T_{\mathrm{MB}}$ map and the observed velocities in the region of the streamer candidate in Fig.~\ref{fig:zoom-hc3n-all} (cont.) center. As there is no recorded velocity for this protostar, we used the \ce{N2H^+} fiber velocity at the location of VLA 3 ($\vlsr=8.4$ \kms) as reference. \draftone{The difference between the \vlsr of the \ce{HC3N} gas and the \ce{N2H^+} \vlsr decreases with increasing distance from VLA 3.} We also plot the outflow directions of SVS 13A and 13B to see if they interfere with the \ce{HC3N} gas that could potentially trace a streamer. The selected \ce{HC3N} component is not affected by the outflow: the redshifted cones of SVS 13A and 13B have $\vlsr > 10$ \kms, whereas the \ce{HC3N} emission from this Gaussian component is lower than 8 \kms; the \ce{HC3N} velocity dispersion is average for its surrounding cloud (0.2 \kms); and finally, there is no discernible velocity gradient in the direction of the outflows (northwest).

Also similar to SK 15, there is a local \ce{N2H^+} peak at the position of the streamer candidate to the north of VLA 3, with two Gaussian components (Fig. \ref{fig:zoom-n2hp-all} middle). Unlike the case of SK 15, the region where there are two Gaussian components is smaller than the area covered by the \ce{HC3N} streamer candidate, barely the size of the \ce{N2H^+} beam. The Gaussian component responsible for the peak is within the cluster that makes up the redshifted fiber and has a velocity of about 8.51 \kms. The second, weakest component shows an almost constant velocity between 7.88 and 7.91 \kms, similar to the velocity of the streamer but without a clear gradient.

\subsubsection{Per-emb 15\label{sec:per15}}

Per-emb-15 \citep[also known as SK 14,][]{Sandell-Knee2001protostars,Enoch2009embeddedprotostars} is a Class I protostar located toward the south of the SVS 13 system. Within the \ce{HC3N} single Gaussian fit, there is a velocity gradient \draftone{within a local} $T_{\mathrm{MB}}$ peak (Fig.~\ref{fig:zoom-hc3n-all} (cont.) right). This structure was not recognized in the HDBSCAN as part of any cluster, possibly because it is located where the fibers overlap. \draftone{An inspection of} the \ce{HC3N} spectra confirmed that there is a well-fitted single Gaussian component along the defined region, and not two Gaussian components fitted as one. The \ce{HC3N} velocity gradient follows the expected direction for asymmetric infall toward the protostellar disk (Fig. \ref{fig:zoom-hc3n-all} (cont.) bottom right). 

Per-emb 15 has no \ce{N2H^+} peak within a beam of its location. the closest \ce{N2H^+} brightness peak is at about 18\arcsec to the northwest, where the outflow traced by MASSES seems to end. The \ce{N2H^+} emission of cluster no. 2 shows a blueshift in velocity toward the protostar. This is more clear in Fig. \ref{fig:zoom-n2hp-all} (cont.) right. We made a PV diagram perpendicular to the outflow direction, shown in Fig.~\ref{fig:zoom-n2hp-all} (cont.) right bottom. 
This diagram is dominated by the emission from the fibers, as this source is located where the two fibers overlap in our line of sight. The blueshifted emission might be produced by the protostar or might be part of the velocity gradient seen in the fiber and caused by gas infall toward the fiber spine.  

\section{Discussion\label{sec:discussion}}

\subsection{Infall of gas onto the fibers}

\draftone{Our results suggest that} \ce{HC3N} and \ce{N2H^+} trace different structures toward NGC 1333 SE. First, both molecules follow the filament direction, but their distributions are different. \ce{N2H^+} recovers more area within the fibers located in this region (Sect. \ref{sec:bayesian}), whereas \ce{HC3N} is more patchy and is \draftone{affected} by the presence of outflows (Sect. \ref{sec:hc3n-outflows}). \draftone{Part of the difference in their structure can be due to the different densities traced, where \ce{HC3N} ($10-9$) molecular line has a critical density about an order of magnitude higher than the \ce{N2H^+} ($1-0$) line, as well as a higher E$_{\mathrm{up}}$ (Table \ref{tab:lines}). \ce{HC3N} has a higher $T_{\mathrm{peak}}$ closer to the protostars. However, we also detected a systematic redshift of \ce{HC3N} with respect to \ce{N2H^+}} in the case of the redshifted fiber, as shown in Sect. \ref{sec:diff-vel}. This result shows that the kinematics of the \ce{HC3N} gas are different from \ce{N2H^+} gas in the redshifted fiber, and thus, \draftone{the \ce{HC3N} is not part of the dense fiber structure but traces dense gas that is less chemically evolved than the gas in the fibers}. 

For the blueshifted fiber, the difference along the line of sight is not significant when compared to the velocity uncertainties (Sect. \ref{sec:diff-vel}). However, it is possible that this is a projection effect and that the motion of \ce{HC3N} with respect to \ce{N2H^+} is mostly perpendicular to our line of sight. Previous works have suggested that gas is being pushed \draftone{by an expanding bubble, responsible for the formation of the filament,}  from the southwest of IRAS 4A \citep{Dhabal2019NGC1333NH3, DeSimone2022shocktrain}. \ce{HC3N} emission traces the blueshifted fiber only toward the west side, which fits into this picture if this newly deposited gas is moved by this expanding shell. We note that close to IRAS 4A and 4B, \ce{HC3N} gas is considerably blueshifted with respect to \ce{N2H^+} gas. This is due to the effect of the outflows of these two protostars that stir the fresh gas more than the dense core structure (traced by \ce{N2H^+}).

\draftone{Given the different kinematics and structures traced, we} propose that the \ce{HC3N} emission represents a layer of chemically fresh material feeding the fibers. In chemical models of low-mass star forming regions, \ce{HC3N} appears at earlier times (a few 100 kyr from $t=0$) than nitrogen-bearing molecules like \ce{N2H^+} \citep{Bergin-Tafalla2007, Sakai-Yamamoto2013WCCCrev}, \draftone{as nitrogen chemistry starts from neutral-neutral reactions, which are much slower than the ion-neutral reactions that dominate the carbon chemistry.} 
\ce{HC3N} has also been suggested as a product of outflow shock fronts \citep{Shimajiri2015outflowchem}, and in our work we observe that outflows help to stir the gas traced in \ce{HC3N}, but we do not include the velocity components of \ce{HC3N} with outflow wings in our clustering (Sect. \ref{sec:clustering}) and subsequent structures' analysis. This ensures that the \ce{HC3N} gas that we compare with \ce{N2H^+} is not affected by possible shock chemical enhancement. This indicates that there is chemical replenishment, delivered from the patchy structure seen in \ce{HC3N}. The fact that \ce{HC3N} is detected in a smaller extension and is patchy with respect to \ce{N2H^+} suggests that the region could be more chemically evolved than other regions in Perseus, e.g. in Barnard 5, where \ce{HC3N} is more extended than the filaments' structures traced by \ce{NH3} \citep{Valdivia-Mena2023B5}. Nevertheless, there is still some fresh material for star formation, in the shape of sparsely distributed gas.

The presence of \ce{N2H^+} perpendicular gradients in both fibers suggests that gas falls toward the fiber spines. These are most prominent toward the south of the filament, where there are no protostars. These kind of gradients are also described for this region in \cite{Dhabal018serpens-perseus} and \cite{Chen2020b-velstructureNGC1333}. \cite{Dhabal018serpens-perseus} suggest the presence of a global velocity gradient for what they call subfilament A, which corresponds in velocity to our redshifted fiber, but it is unclear for their subfilament B, corresponding approximately in \vlsr with our blueshifted fiber. We confirmed that there is also a perpendicular velocity gradient for the blueshifted fiber in \ce{N2H^+}. As suggested in these previous works, the velocity gradients observed in the fibers can be related to accretion flow toward the fibers, seen as contraction of a sheet-like cloud \citep{Chen2020filamentform-simulations}. The velocity gradients could also be consistent with fiber rotation, but in simulations of filament formation, these motions do not seem favorable in fibers \citep{Smith2016filformsim}. \ce{HC3N} also shows prominent perpendicular velocity gradients where there are no protostars, but the regions which follow this motion are smaller. In general, the gradients in \ce{HC3N} tend to change more drastically near protostars (Fig. \ref{fig:HC3N-N2Hp-subfilaments-clusters} \refereeone{top}). This suggests that the presence of the protostars stirs the surrounding gas, which is reflected in more complex gas motions at smaller scales.

\subsection{Discovery of streamer \draftone{candidates} in NGC 1333 SE}

We found streamer candidates toward 7 out of 16 \refereeone{YSOs} within our field of view. 
Moreover, we confirmed the infalling motion toward one of them (IRAS 4A). For the rest \draftone{of the candidates}, we do not have enough \draftone{spatial} resolution or information about the protostar and disk masses to efficiently model the free-falling gas. \draftone{What we can say about their three-dimensional structure is that these streamers come from behind the protostars, as they all become more blueshifted with respect to the protostellar \vlsr with decreasing distance.} The streamer candidates in our work were found towards Class 0, 0/I and I protostars. This represents a streamer frequency of approximately 40\% if we consider all the \refereeone{YSOs} within the maps' field of view. If we consider only the early stage protostars (Class 0 to I, for which we have \draftone{12} within the map), the frequency of streamers is higher, about 60\%. Although these are small number statistics, it is a first approach to quantify the prevalence of streamers toward \refereeone{YSOs} \draftone{(if we are able to confirm the infall nature of the emission in the future)}. 

This is a \refereeone{first coarse estimate} to the real frequency of streamers in the area, as \refereeone{these are candidates still not confirmed dynamically, and} it is possible we have missed streamers. First, we \refereeone{are unable to observe} any streamer that is smaller than \draftone{approximately} 1500 au in projected length, as that is our resolution limit. \draftone{This decreases our chances to find small streamers and also those almost fully contained along our line of sight.} In particular, \draftone{this limits our chances to find Class II streamers (of which we do not find any), as they} are usually on the order of 300 to 500 au \citep{Garufi2021accretionDGTauHLTau, Ginski2021}, although longer arms of up to $\sim2000$ au surrounding T Tauri disks have been suggested to be streamers \citep{Alves2020,Huang2021MAPSstreamersGMAur, Huang2022arcDOTau}. 
Most Class II sources in the field have no detected \ce{HC3N} emission: only one Class II source (ASR 54) has \ce{HC3N} emission detected around it, and its velocity structure is not suggestive of streamer motion, but \draftone{appears} dominated by the infall of fresh gas into the fiber. This does not mean that there are no streamers toward these sources, but the tracer (\ce{HC3N}) and/or the resolution are not adequate to find streamers toward these sources. 
Secondly, the assumption of \draftone{monotonically decreasing velocity with distance} along the line of sight \draftone{can} miss streamers. This assumption \draftone{helps} to recognize infalling motion and to differentiate it from outflowing gas, \draftone{but} there are instances where \draftone{a streamer} will not show this kind of gradient along the line of sight. If the acceleration is completely contained in the plane of the sky, the observed \vlsr of the gas is constant. \draftone{It is also possible that, due to projection effects, the velocity appears to be monotonically \textit{increasing} \refereeone{in the outer regions of the streamer}, such as in the case of Per-emb 2 \citep{Pineda2020Per2}}, confirmed using the free-fall analytic solution from \cite{Mendoza2009}. As an acceleration proportional to distance is usually attributed to outflowing gas, it is necessary to take other factors into account (like the shape of the emission and the position of the known outflow) to disentangle outflowing emission from inflowing gas. \draftone{Considering the spatial and spectral resolution of our \ce{HC3N} data and the median velocity gradient found in our streamers ($\sim 15$ \kms pc$^{-1}$), the probability to miss a streamer is about 50\%, either because it is contained along our line of sight or close enough to the plane of the sky that its gradient is not observable.}

\draftone{We define the structures found in this work, including toward IRAS 4A, as streamer candidates}. 
The limiting factors at the time of describing the structure of the streamers is the lack of angular resolution \draftone{and the lack of information about the protostar and disk masses}. We show that \ce{HC3N} traces infall given the velocity of the gas that surrounds a protostar and the location of the local velocity gradient, but we do not have enough information \draftone{about the protostars themselves} to confirm that the velocity profiles correspond to free-fall. To confirm the streamer nature of these candidates, it is necessary to replicate their structure (a thin and long structure in $T_{\mathrm{MB}}$ and a velocity gradient) with a free-fall model. In the case of IRAS 4A, even if we could model the infall, the resolution \draftone{is not high enough} to describe the structure in the image plane. 
Therefore, we require follow-up observations with a resolution higher than 4.9\arcsec to fully describe the structure of these streamers.

\subsection{Relation of streamers to the larger gas infall}

Our results indicate that the gas that builds up the streamers \draftone{comes from beyond the fibers.} There are \draftone{three} main reasons for this conclusion. First, we did not \refereeone{observe} signs of infall using \ce{N2H^+}. The \ce{N2H^+} \vlsr close to each protostar is different than the \ce{HC3N} \vlsr, and we do not observe stream-like structures in \ce{N2H^+}, \draftone{although this could be due to projection effects}. Second, by construction, the structure of the fibers is traced using \ce{N2H^+}, and we observed significant differences in the central velocities of the redshifted fiber with respect to the \ce{HC3N} flows within. In the case of the blueshifted fiber, we did not see a significant difference (Sect. \ref{sec:diff-vel}). However, this might be a projection effect, where we did not see signs of any kinematic difference because the movement is along the plane of the sky. \draftone{Third, the apparent direction of the streamer candidates does not coincide with the orientation of the filament.} 

\draftone{Our} results suggest that the mass that composes the streamers does not come from the fiber structure, but from the fresh gas that is infalling toward the fibers. Although \ce{HC3N} can be enhanced by the presence of outflows, we took out the velocity components affected by outflows for this analysis, so the streamer gas being chemically fresh is the most probable origin. A similar relation between gas outside the filament and streamers is suggested for Barnard 5, another region in Perseus, but with different tracer for each scale \citep{Valdivia-Mena2023B5}. This result is consistent with research that indicates that cores must be replenished with fresh gas to form protostars, as the amount of envelope mass dispersed through the outflow is substantial \citep{Hsieh2023outflowcavities}. Streamers, therefore, can be the mechanism that can feed the protostellar system with mass that is not from the original core, connecting the material coming from outside the fibers and delivering it toward the protoplanetary disks. 

\section{Conclusions\label{sec:conclusions}}

In this work, we analyzed the distribution and velocity structure of the NGC 1333 SE fibers in search for accretion signatures toward \refereeone{YSOs}. Our results are summarized below.

\begin{itemize}
    \item The distribution of \ce{HC3N} gas in this region is patchy and does not cover the full extent of the fibers seen in \ce{N2H^+}. The \ce{HC3N} velocity along the line of sight is redshifted with respect to \ce{N2H^+} in the redshifted fiber. \draftone{Together, these results} indicate that \ce{HC3N} is following different kinematics than \ce{N2H^+}. We suggest \ce{HC3N} traces gas that is infalling later to the filament, after star formation has started in the region. 
    \item \ce{N2H^+} shows velocity gradients perpendicular to the fibers' orientations. We suggests, as previous works have done as well \citep{Chen2020b-velstructureNGC1333, Dhabal018serpens-perseus}, that this indicates infall toward the fibers's spines.
    \item The outflows of IRAS 4A and IRAS 2A generate wings and strong brightness peaks in the \ce{HC3N} spectra. In these regions, \ce{HC3N} is enhanced at the bow shocks of the protostellar outflows. 
    \item We found streamer candidates toward 7 out of \draftone{16} \refereeone{YSOs} in the field of view of our mosaic. \textbf{This represents an \draftone{incidence} of about \draftone{40\%} of \refereeone{YSOs} with streamers when looking within a \draftone{region}.} The 7 candidates are all found toward early stage protostars (Class 0 to I), which represent a total of 12 sources within our field of view, so for early stages in particular, the \draftone{incidence} of streamers is about 60\% for this given filament.
    \item The gas that composes the streamers is coming from outside the fiber structure, as there is a difference between the velocity structure of \ce{HC3N} with \ce{N2H^+} and no streamers are detected in \ce{N2H^+} emission. Only 2 protostars show \ce{N2H^+} emission with similar velocities in the same position as the \ce{HC3N} streamers, but do not show the same velocity profiles.
\end{itemize}

We defined these \draftone{structures as candidate streamers} because the resolution of our data is not enough to resolve the width of the flow and does not allow for an accurate modelling of the infall. \draftone{Further information on the mass and orientation of the protostars and disks in the region} are required to model the infall and determine the true length of the streamers. If we take NGC 1333 as a ``typical" star-forming region, then we expect streamers to be a frequent feature toward protostellar envelopes. This work highlights the relevance of streamers in our new picture of low-mass star formation.

\begin{acknowledgements}
	\refereeone{The authors wish to thank A. Hacar and M. Chen for sharing their original datasets with us for our analysis, and the anonymous referee for their careful review of the manuscript.}
M.T.V., J.E.P., P.C., S.S., A.I. and M.J.M. acknowledge the support by the Max Planck Society. 
D. S.-C. is supported by an NSF Astronomy and Astrophysics Postdoctoral Fellowship under award AST-2102405. SO was supported by NSF AAG 2107942.
M.K. acknowledges funding from the European Union’s Framework Programme for Research and Innovation Horizon 2020 (2014-2020) under the Marie Skłodowska-Curie Grant Agreement No. 897524 and funding from the Carlsberg foundation (grant number: CF22-1014). 
\end{acknowledgements}

\bibliographystyle{aa} 
\bibliography{main} 

\appendix

\section{Channel maps\label{ap:chanmaps}}

\begin{figure*}
    \centering
    \includegraphics[width=0.98\textwidth]{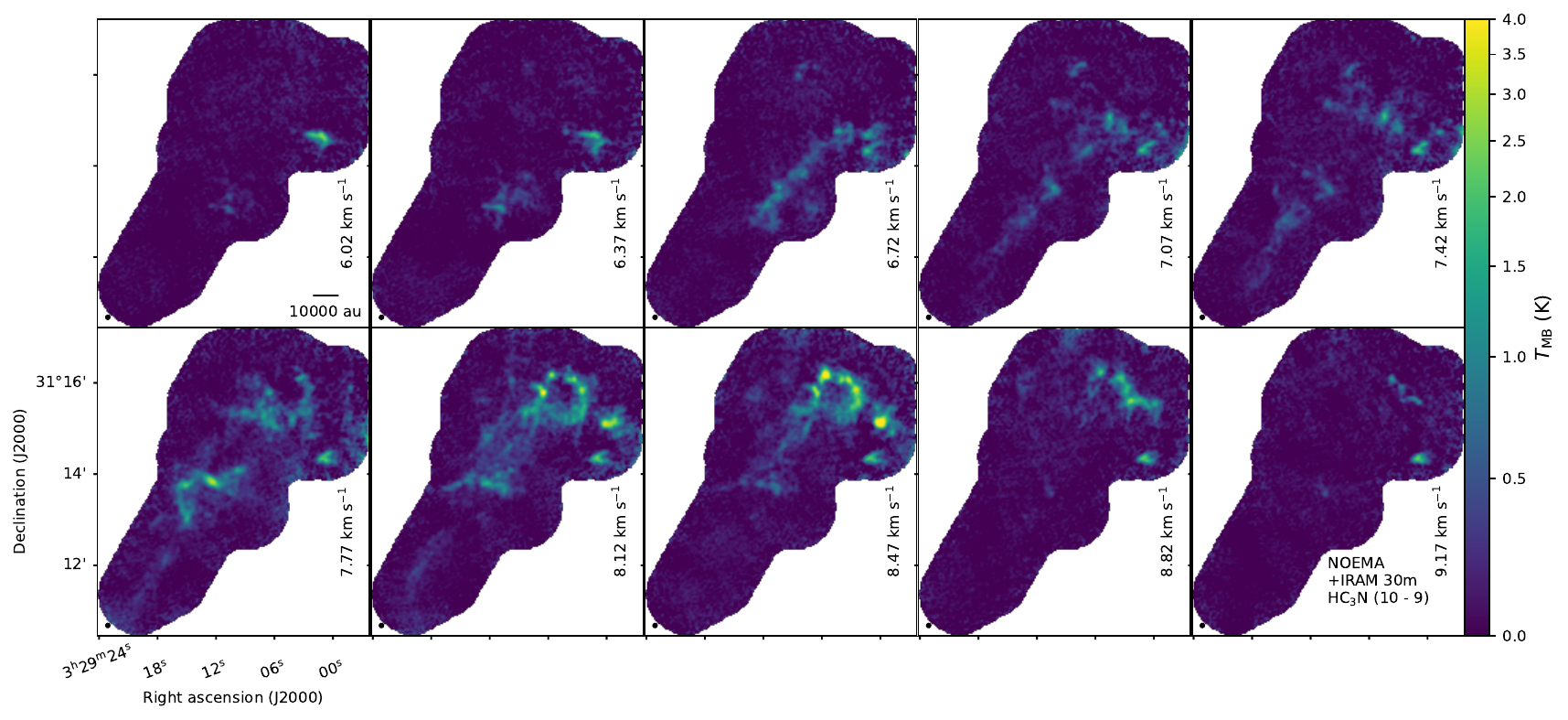}
    \caption{Channel maps between \refereeone{6.02} and \refereeone{9.17} \kms for \ce{HC3N} $J=10-9$ emission. }
    \label{fig:HC3N-chanmap}
\end{figure*}

\begin{figure*}
    \centering
    \includegraphics[width=0.98\textwidth]{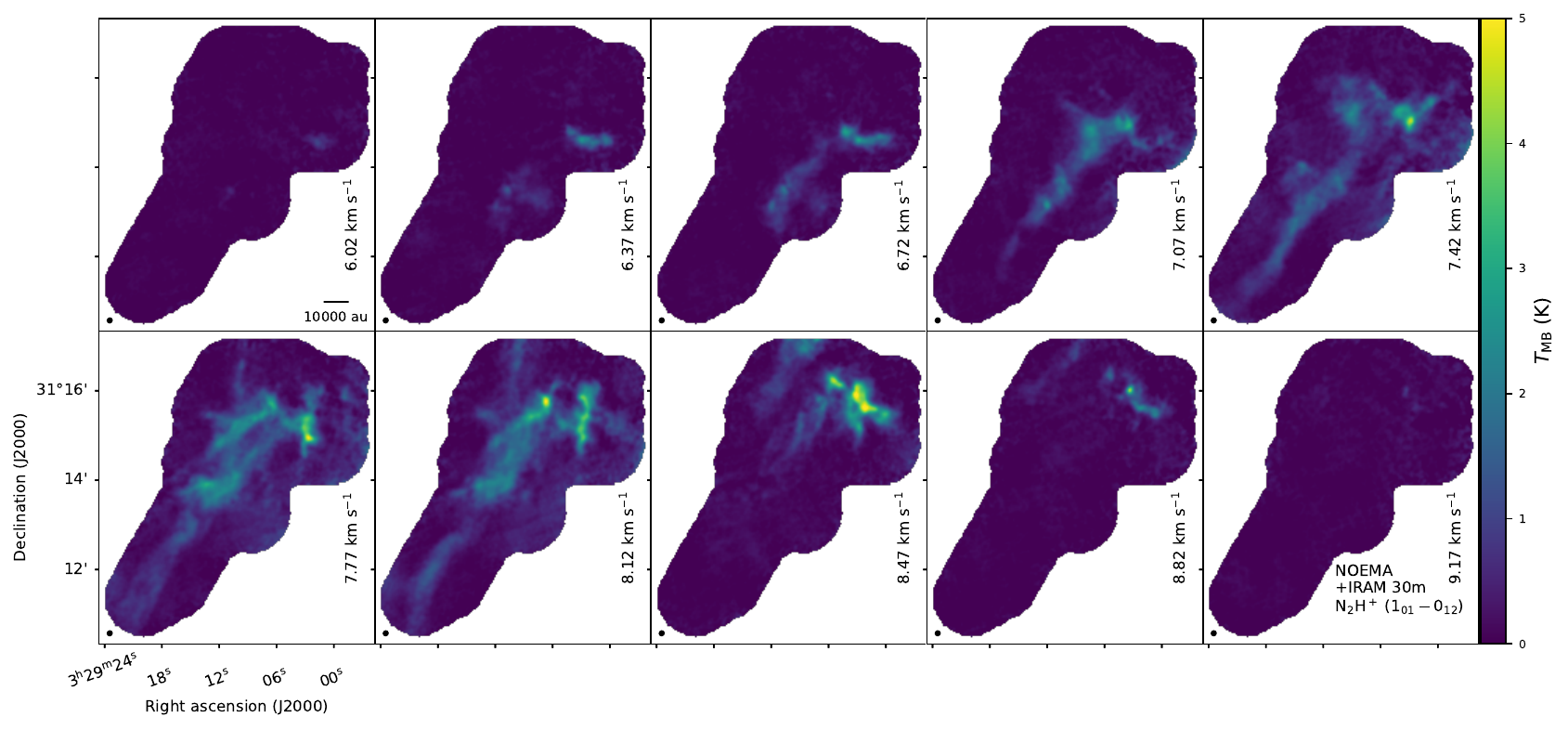}
    \caption{Channel maps between \refereeone{6.02} and \refereeone{9.17} \kms for \ce{N2H^+} $J=1-0$, $F_1F=01-12$  emission. }
    \label{fig:N2Hp-chanmap}
\end{figure*}

Figures \ref{fig:HC3N-chanmap} and \ref{fig:N2Hp-chanmap} show the channel maps of \ce{HC3N} and \ce{N2H^+} emission between approximately 6 and 9.2 \kms, in steps of about \refereeone{0.35} \kms. We note that there is emission in \ce{HC3N} outside of this range, starting from -3 \kms and ending at approximately 13 \kms. 

\section{Density-based clustering of molecular emission\label{ap:clustering}}

In this section, we describe the steps taken to cluster the \ce{N2H^+} and \ce{HC3N} velocity structures. We cluster the Gaussian points from each molecule based on their position in the plane of the sky and their $v_{\mathrm{LSR}}$. Using the amplitude $T_{\mathrm{peak}}$ or their dispersion  $\sigma_{\mathrm{v}}$ does not improve the clustering results, and as our goal is to disentangle the velocity structures of the filament, the analysis in the position-position-velocity space is sufficient (in particular after filtering out the Gaussians with high dispersion). 

From the Gaussian fitting and the quality assessment, we have 94864 Gaussians in the \ce{N2H^+} data cube and 56771 Gaussians in the \ce{HC3N} data cube. We did a finer selection of Gaussian components for both molecules before clustering, so as to ensure that we trace the bulk emission of the filament structure. We select only the points with low uncertainty in their parameters, that is, with an uncertainty less than 25\% the value of the parameter. This criterion leaves out uncertain points that can add confusion to the clustering. We also filter out the Gaussian fits that show $\sigma_{\mathrm{v}}>1$ \kms, as these represent fits with outflows. We filter possible Gaussian components that represent random noise in the spectra by selecting only those components with velocities between 5 and 9 \kms. This also helps to filter out \ce{HC3N} that corresponds to high-velocity outflow wings. We were left with 81657 Gaussians in the \ce{N2H^+} data cube and 52019 Gaussians in the \ce{HC3N} data cube.

\draftone{We cluster the Gaussian component results of \ce{N2H^+} using Hierarchical Density-Based Spatial Clustering of Applications with Noise (HDBSCAN). HDBSCAN is an extension of the Density-Based Spatial Clustering of Applications with Noise (DBSCAN) algorithm. 
In summary, DBSCAN defines clusters of points in the user-defined hyperspace as local point overdensities, and leaves sparsely distributed points as noise \citep{Ester1996DBSCAN}. A core is defined as an overdensity surrounding a core sample (a single point in the hyperspace) with a minimum of samples $n$ within a radius of $\epsilon$. HDBSCAN, instead of fixing the radius $\epsilon$, selects clusters based on the minimum spanning tree of the mutual reachability graph\footnote{https://scikit-learn.org/stable/modules/clustering.html}, i.e. explores all possible values of $\epsilon$ \citep{Campello2013HDBSCAN, McInnes2017HDBSCAN}. This process allows the algorithm to form clusters of different densities. We use the \texttt{hdbscan} package from the contributed packages to \texttt{scikit-learn}\footnote{https://hdbscan.readthedocs.io/en/latest/}. }

\begin{table}[ht]
	\centering
	\caption{\label{tab:paramsclustering}HDBSCAN parameters used to cluster the \ce{HC3N} and  \ce{N2H^+} Gaussian peaks.}
\begin{tabular}{lll}
	\hline\hline
	Parameter  & \ce{HC3N}  & \ce{N2H^+}     \\ \hline
	min\_sample & 200   & 240 \\
	min\_cluster\_size  & 500    & 900  \\ \hline
\end{tabular}

\end{table}

\draftone{The best results for each molecule are obtained with the parameters shown in Table \ref{tab:paramsclustering}, named according to the parameter names in the python implementation\footnote{https://hdbscan.readthedocs.io/en/latest/parameter\_selection.html}. The parameters are different for each molecule because the number of Gaussian peaks for  \ce{HC3N} (about 51000) is much smaller than the number of Gaussian peaks in \ce{N2H^+} (more than 81000). The resulting clusters can be seen in Fig.~\ref{fig:cluster-results}. }

\begin{figure*}
    \centering
    \includegraphics[width=0.9\textwidth]{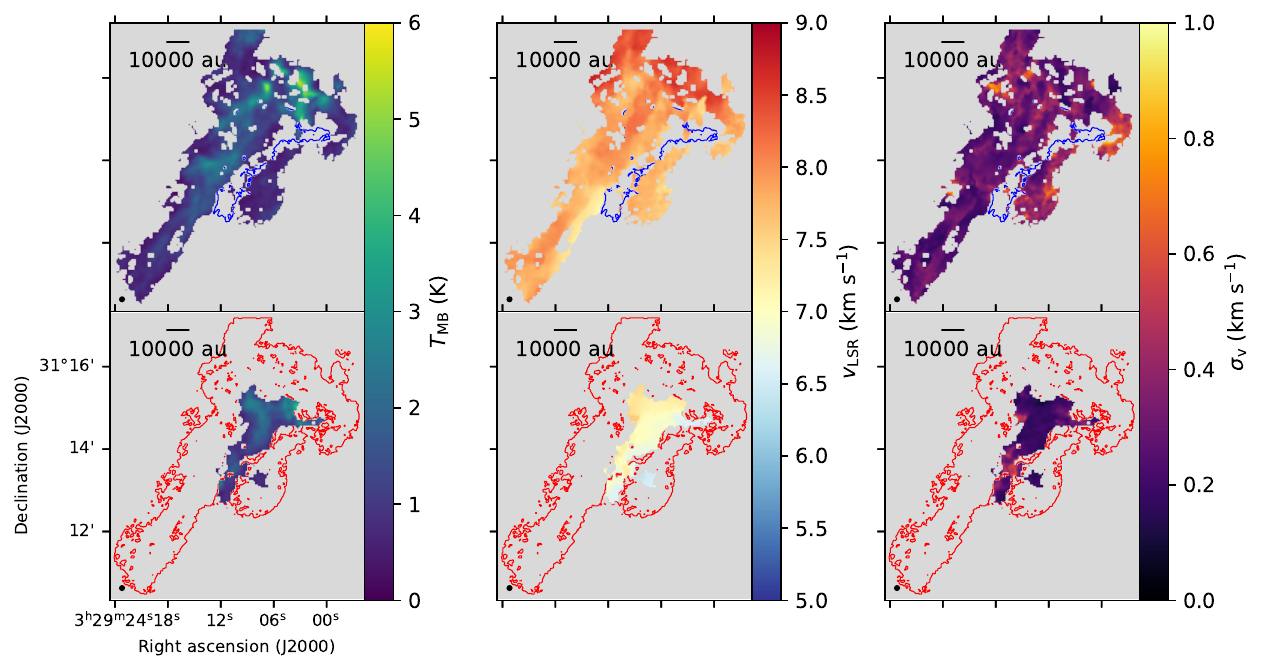}
    \caption{Peak temperature, central velocity and velocity dispersion of the \ce{N2H^+} clusters. Top: $T_{\mathrm{peak}}$, \vlsr and $\sigma_{\mathrm{v}}$ for the redshifted fiber. The blue contour indicates the position of the blueshifted fiber. Bottom:  $T_{\mathrm{peak}}$, \vlsr and $\sigma_{\mathrm{v}}$ for the blueshifted fiber. The red contour indicates the position of the redshifted fiber. }
    \label{fig:N2Hp-clusterprops}
\end{figure*}

\begin{figure*}
	\centering
	\includegraphics[width=0.9\textwidth]{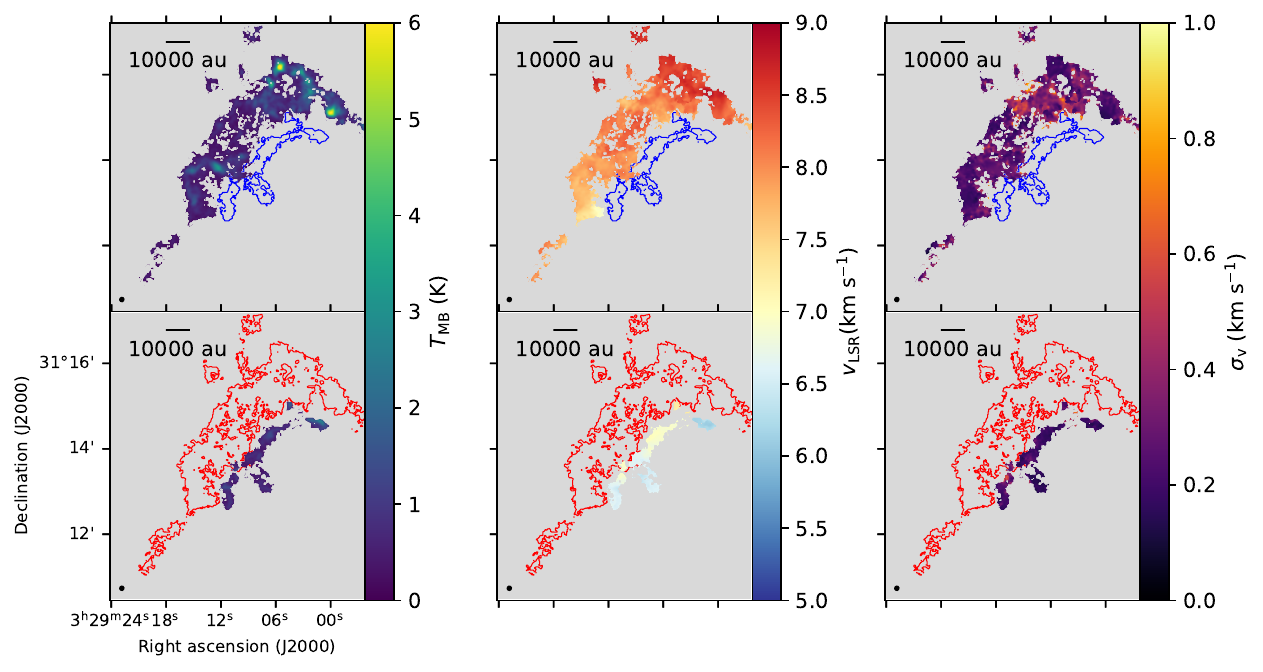}
	\caption{Same as Fig. \ref{fig:N2Hp-clusterprops} but for the \ce{HC3N} clusters. }
	\label{fig:HC3N-clusterprops}
\end{figure*}

\refereeone{Figure \ref{fig:N2Hp-clusterprops}} shows the peak temperature $T_{\mathrm{peak}}$, velocity \vlsr and dispersion $\sigma_{\mathrm{v}}$ of the \ce{N2H^+} clusters, grouped according to their fiber, redshifted or blueshifted. resulting from the HDBSCAN analysis. 
\refereeone{Figure \ref{fig:HC3N-clusterprops}} shows the same quantities but for the \ce{N2H^+} clusters.

\section{\refereeone{Comparison of clustering results with previous works\label{ap:fiber-comparison}}}

\begin{figure*}
	\centering
	\includegraphics[width=0.8\textwidth]{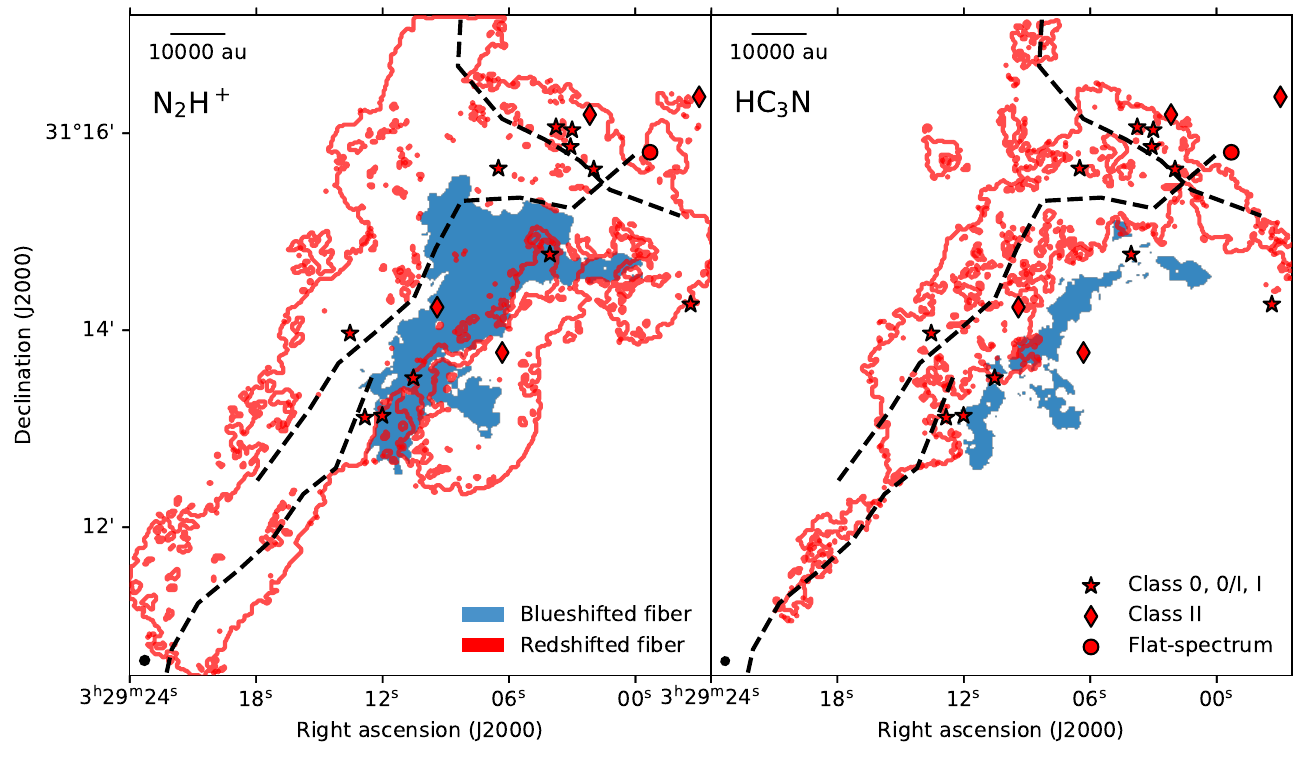}
	\caption{\refereeone{Clusters corresponding to the redshifted and blueshifted fibers in \ce{N2H^+} and \ce{HC3N} (as Fig. \ref{fig:cluster-results} left), with the \ce{N2H^+} fiber spines taken from \cite{Hacar2017NGC1333}, shown as dashed black lines.} }
	\label{fig:cluster-fiber-comparison-ap}
\end{figure*}

\refereeone{Figure \ref{fig:cluster-results} show that the positions of the clusters found for \ce{N2H^+} emission are consistent with the \ce{NH3} fibers found in \cite{Chen2020b-velstructureNGC1333}. However, our \ce{N2H^+} clustering is different than the fiber structure found by \cite{Hacar2017NGC1333}, using the same molecule and hyperfine transition. Figure \ref{fig:cluster-fiber-comparison-ap} left shows the \ce{N2H^+} fiber spines from \cite{Hacar2017NGC1333} over the clusters found in \ce{N2H^+} emission. Our clustering does not recognize a third separate fiber toward the SVS 13 region as in the case of \cite{Hacar2017NGC1333}.  The redshifted cluster found in our work coincides well with the leftmost fiber from \cite{Hacar2017NGC1333} \citep[as well as the eastern fiber from][located at the same position]{Chen2020b-velstructureNGC1333}. The blueshifted cluster, on the other hand, does not coincide with the rightmost fiber from \cite{Hacar2017NGC1333}, which connects better to the southernmost part of our redshifted cluster. This mismatch between structures found using the same molecule and transition is due to the different algorithms used for structure recognition and the differences in resolution between the datasets. }

\refereeone{We do not expect to obtain the same fiber structures as previous works as we use a different, more general algorithm. HDBSCAN is a purely mathematical algorithm with no physical information. Our clustering is based on proximity in a scaled PPV space, without considering other parameters such as brightness or line width, whereas Friends in Velocity (FIVE) is designed to identify velocity-coherent structures based on emission intensity and velocity, considering the PPV space is Nyquist-sampled and represents a fluid \citep{Hacar2013FIVE}. }

\refereeone{The different resolutions contribute to the different structures recognized by the different algorithms. The beam of our data ($\sim 6\arcsec$, Table \ref{tab:lines}) is about 5 times smaller than for the data used in \cite{Hacar2017NGC1333} (30 \arcsec), so our data resolves the structure \textit{within} the fibers identified in previous works. When the beam increases, the brightness sensitivity is better, but the power to resolve substructures decreases, so even though the emission that corresponds to our blueshifted fiber might have been detected with IRAM 30-m, it was not large enough to categorize in its own structure and was left as noise by FIVE. Our clustering is used as a guide to match which velocity components from different molecules can be studied together, so as to understand the difference in kinematics between \ce{N2H^+} and \ce{HC3N} (Sect. \ref{sec:diff-vel}). Also, at this scale, fiber-identification algorithms such as FIVE will probably find further substructures. Future works will explore what kind of substructures can be found using fiber-identification algorithms in high-resolution observations. The comparison of our clustering results with their recovered fibers shows that structure identification in PPV data is dependent on both the data itself and the algorithm chosen to identify structure.}

\section{Close-up of \ce{N2H^+} velocity profiles toward individual protostars}

Figure \ref{fig:zoom-n2hp-all} shows the same two top panels ($T_{\mathrm{peak}}$ and \vlsr) \refereeone{of Fig. \ref{fig:zoom-iras4a-hc3n-n2hp} right} for \refereeone{all other} sources where we find a streamer candidate. Some show position-velocity cuts, and for those where a second component in \ce{N2H^+} coincides roughly with the streamer's velocity, we plot the \vlsr KDE of the corresponding component instead of the position-velocity cut. 

\begin{figure*}
    \centering
    \includegraphics[width=0.7\textwidth]{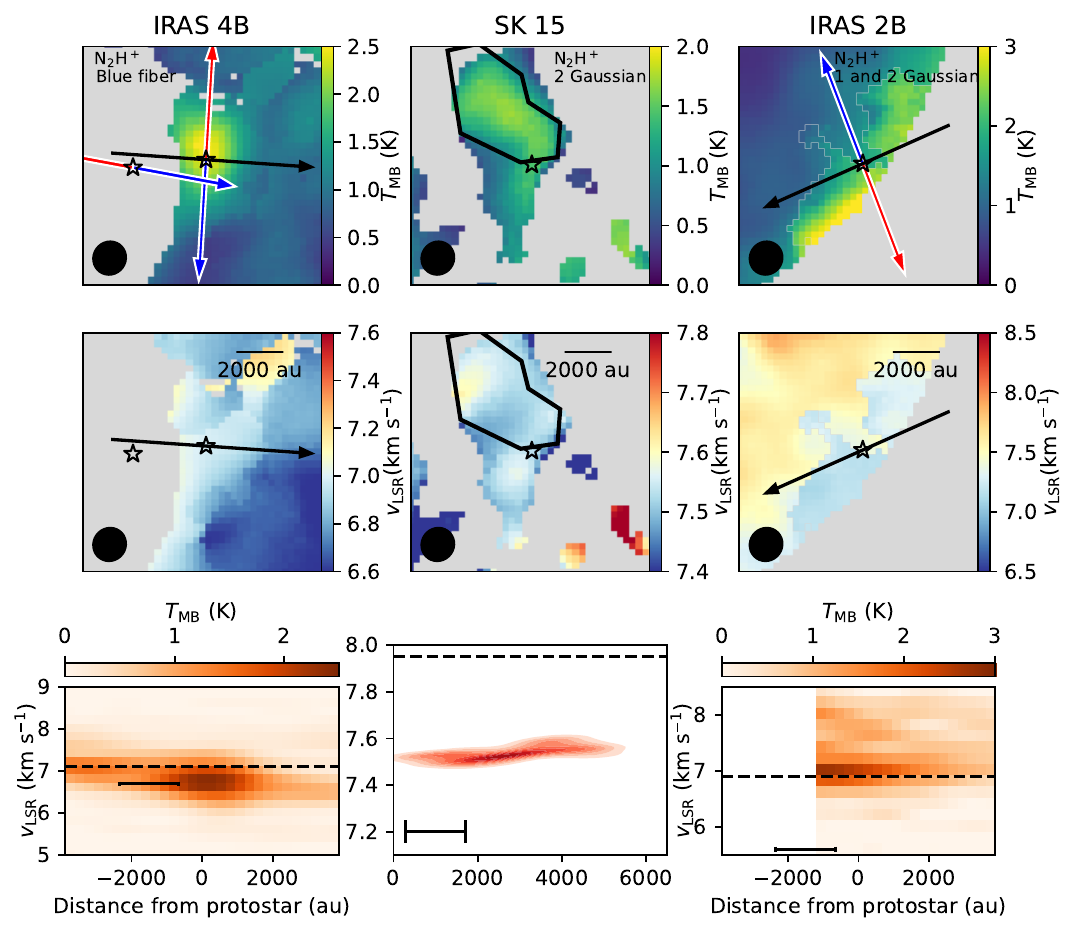}
    \caption{Zoom-in plots of \ce{N2H^+} emission for IRAS 4B (left), SK 15 (center) and IRAS 2B (right), with the same panels as shown in Fig. \ref{fig:zoom-iras4a-hc3n-n2hp} \refereeone{right} except for SK 15. The Gaussian component for each protostar is labeled accordingly. The black ellipse at the bottom left corner represents the beam. Top: Amplitude $T_{\mathrm{MB}}$ of the Gaussian component plotted. The blue and red arrows indicate the direction of the blueshifted and redshifted outflow lobes, respectively, for known outflows in the plotted area.  Middle: Central velocity \vlsr of the Gaussian component selected. The scalebar represents a length of 2000 au. Bottom: For IRAS 4B and IRAs 2B, the position-velocity diagram along the path indicated in the top panel. For SK 15, KDE of the \vlsr within the selected region. The red density histogram representd the KDE of the velocities within the black polygon. The dashed lines mark the \vlsr of each protostar. The black scalebar represents a length equivalent to one beam.  }
    \label{fig:zoom-n2hp-all}
\end{figure*}

\begin{figure*}
    \ContinuedFloat
    \captionsetup{list=off,format=cont}
    \centering
    \includegraphics[width=0.7\textwidth]{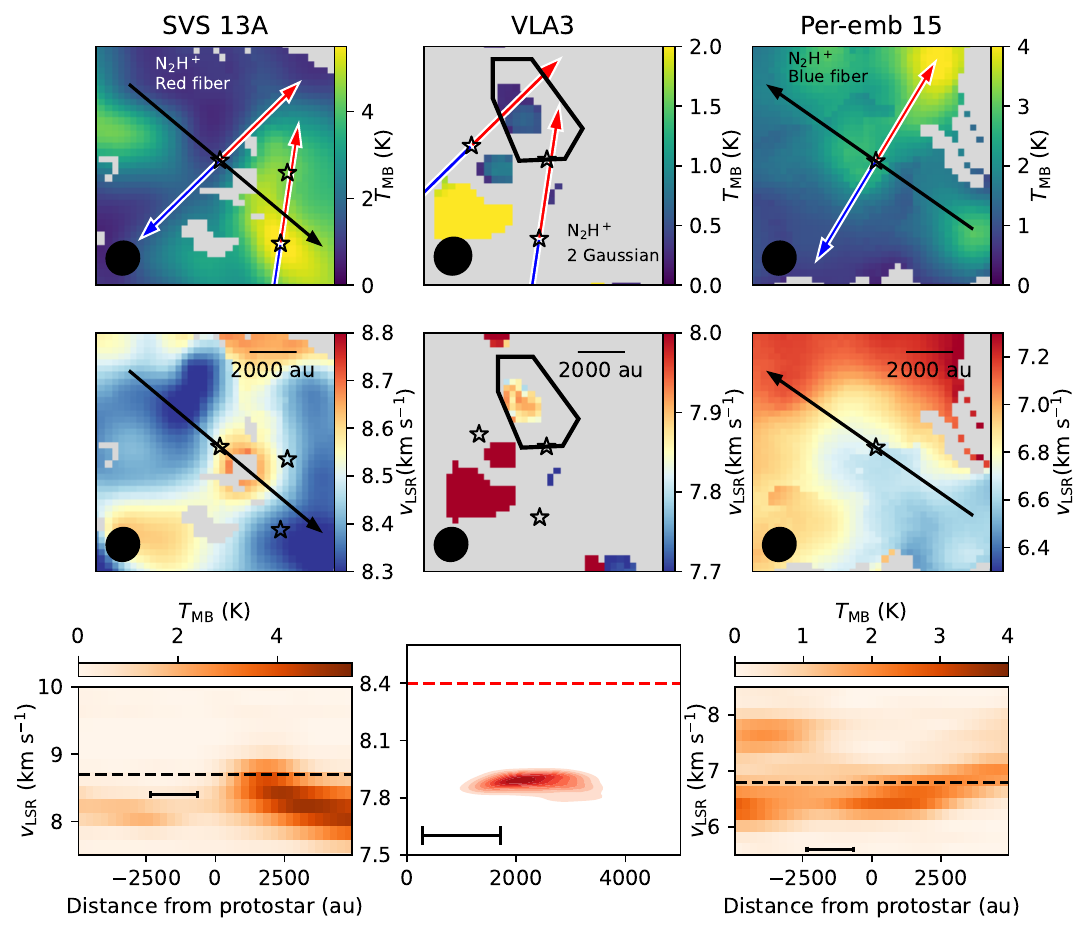}
    \caption{Zoom-in plots of \ce{N2H^+} emission for SVS 13A (left), VLA 3 (center) and Per-emb 15 (right), with the same panels as shown in Fig. \ref{fig:zoom-iras4a-hc3n-n2hp}, except for VLA 3. }
\end{figure*}

\end{document}